\documentclass[lettersize,journal]{IEEEtran}
\usepackage{amsmath,amsfonts}
\usepackage{algorithmic}
\usepackage{algorithm}
\usepackage{array}
\usepackage[caption=false,font=normalsize,labelfont=sf,textfont=sf]{subfig}
\usepackage{textcomp}
\usepackage{stfloats}
\usepackage{url}
\usepackage{verbatim}
\usepackage{graphicx}
\usepackage{cite}
\usepackage{booktabs} 
\begin{document}

\title{Cross-Subject Generalization in Decoding Perceived Speech from Non-Invasive Brain Recordings}

\author{Aoke Zhang, Bo Wang, Xihong Wu, \textit{Senior Member, IEEE}, Heping Cheng, and Jing Chen

\thanks{This work was supported by the STI 2030-Major Projects (Grant No.2021ZD0201500). (First author: Aoke Zhang.) (Corresponding author: Jing Chen.)}

\thanks{Aoke Zhang is with the Speech and Hearing Research Center, School of Intelligence Science and Technology and the Center for BioMed-X Research, Academy for Advanced Interdisciplinary Studies, Peking University, Beijing, China.}

\thanks{Bo Wang and Xihong Wu are with the Speech and Hearing Research Center, School of Intelligence Science and Technology, Peking University, Beijing, China, and also with the National Key Laboratory of General Artificial Intelligence, Beijing, China.}

\thanks{Heping Cheng is with the Center for BioMed-X Research, Academy for Advanced Interdisciplinary Studies, Peking University, Beijing, China.}

\thanks{Jing Chen is with the Speech and Hearing Research Center, School of Intelligence Science and Technology, Peking University, Beijing, China, and also with the Center for BioMed-X Research, Academy for Advanced Interdisciplinary Studies, Peking University, Beijing, China, and also with the National Key Laboratory of General Artificial Intelligence, Beijing, China. (e-mail: janechenjing@pku.edu.cn)}
}

\markboth{IEEE/ACM TRANSACTIONS ON AUDIO, SPEECH AND LANGUAGE PROCESSING, VOL. XX, NO. XX, XX XXXX}%
{Shell \MakeLowercase{\textit{et al.}}: A Sample Article Using IEEEtran.cls for IEEE Journals}


\maketitle

\begin{abstract}
Decoding perceived speech from non-invasive brain recordings has garnered significant attention in recent years due to its wide range of potential applications. However, existing methods face considerable challenges in cross-subject decoding, primarily due to limited generalizability and the absence of explicit mechanisms for extracting subject-consistent information. These limitations result in high training costs and suboptimal decoding performance. To address these challenges, we propose an innovative Cross-Subject Perceived Speech Decoding (CPSD) framework, which comprises two training stages: source model pre-training and personal specialization. In the source model pre-training stage, contrastive learning is employed to capture shared representations across multiple source subjects. Subsequently, personal specialization initializes the model for the target subject by extracting consistent components from the source model and fine-tuning it using target subject data. Additionally, we introduce the Positional Encoding-based Spatial Attention (PESA) module, which remaps MEG/EEG data into a standardized reference space, thereby enhancing cross-subject consistency and facilitating model training. We evaluate the proposed CPSD framework on three perceived speech neural datasets encompassing different modalities and languages. The results demonstrate that our framework outperforms baseline methods by more than 6.8\%, 15.4\%, and 15.8\% in Top-10 accuracy on the Armeni 2022, PKUEEG 2025, and Broderick 2018 datasets, respectively. Further analyses confirm the effectiveness, efficiency, and robustness of the proposed approach. 
\end{abstract}


\section{Introduction}

\begin{figure}[t]
\centering
\includegraphics[width=8.8cm, height=6.72cm]{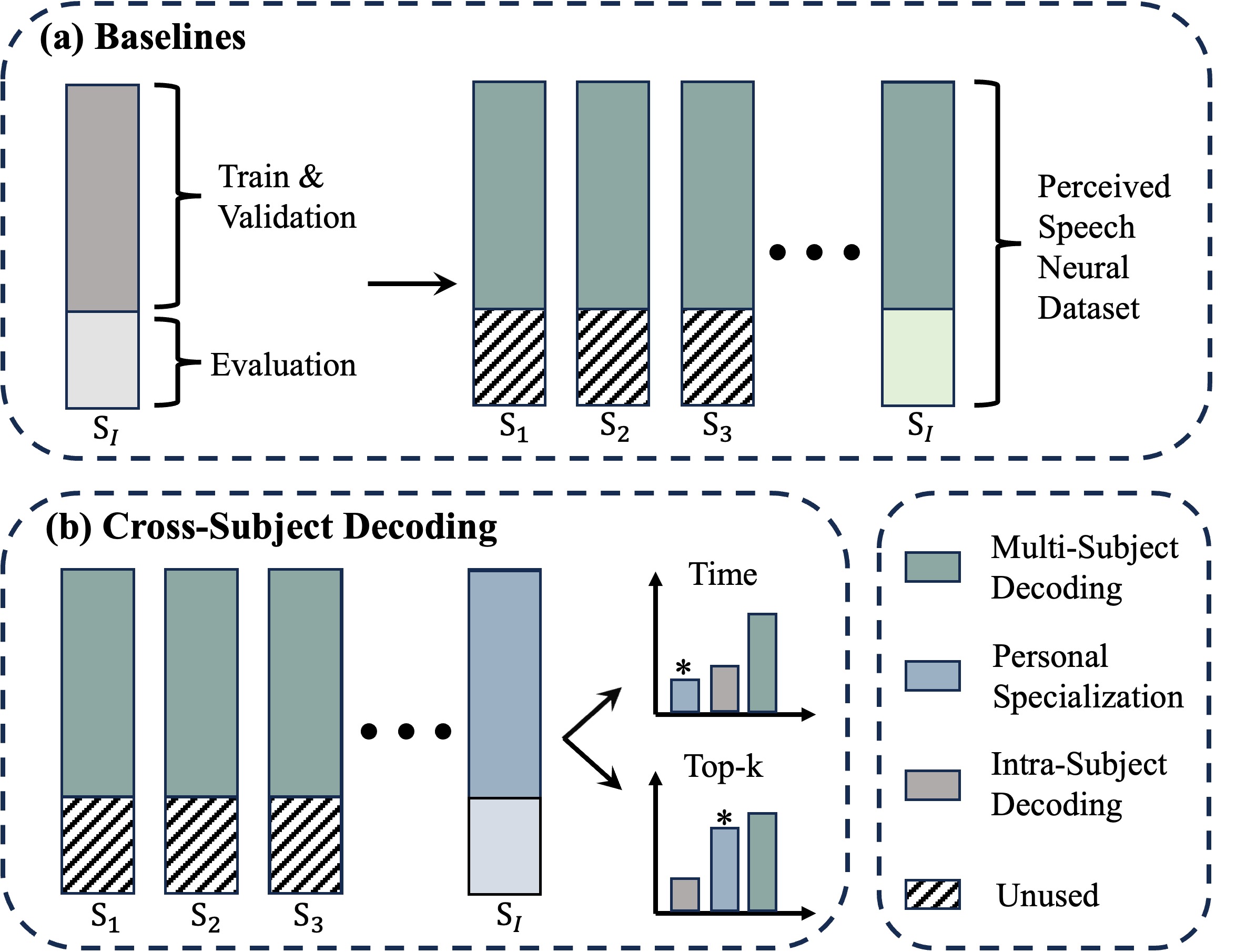}
\caption{Different settings for perceived speech decoding include: (a) intra-subject decoding, which involves decoding exclusively within the target subject; multi-subject decoding, where the model is trained on data from all subjects simultaneously and then evaluated on the target subject; and (b) cross-subject decoding, which entails pre-training the model on source subjects followed by generalization and evaluation on the target subject.}
\label{definition}
\end{figure}

\IEEEPARstart{N}{eural} speech decoding, which directly transforms neural signals into speech, has made significant progress in recent years \cite{levy2026noninvasive,wairagkar2025instantaneous,card2024accurate,defossez2023decoding,tang2023semantic}. Among the various paradigms of neural speech decoding, decoding perceived speech has been extensively studied due to its importance in understanding the mechanisms of speech processing \cite{bhattacharjee2026aligning,zou2026constituent,tang2023semantic,akbari2019towards}, the large volume of available data, and the strong alignment between neural signals and speech stimuli. Furthermore, because of the overlap between the neural mechanisms underlying speech perception and speech production \cite{hickok2007cortical,skipper2005listening,kraemer2005sound,wheeler2000memory}, related research holds potential to advance the development of speech neuroprostheses \cite{wairagkar2025instantaneous,card2024accurate,silva2024speech}, which can restore communication for patients who have lost the ability to speak. Invasive brain-computer interfaces (BCIs) provide high-quality recordings of speech perception processes. However, they require surgical implantation, which limits accessibility for typical users \cite{leuthardt2021defining}. Additionally, due to immune rejection reactions, maintaining signal quality over the long term is challenging \cite{chen2025long, yang2025neural, polikov2005response}. Moreover, the coverage of invasive BCI-implanted electrode arrays is limited. For example, in speech perception research, electrode arrays can be implanted near the superior temporal gyrus and middle temporal gyrus \cite{akbari2019towards}. In contrast, magnetoencephalography (MEG) and electroencephalography (EEG) are widely used non-invasive BCI techniques that offer greater safety, high temporal resolution, and whole-brain recording capabilities \cite{peksa2023state}.  Consequently, a considerable amount of research has focused on this area \cite{zhang2025novel,wang2024semantic,defossez2023decoding}. The typical task of decoding perceived speech from non-invasive brain recordings involves retrieving the specific speech stimulus from test samples to which a subject is listening \cite{wang2026hierarchical,zhang2025novel,defossez2023decoding}.

Decoding speech from MEG/EEG signals remains challenging due to a low signal-to-noise ratio (SNR), limited spatial resolution \cite{sanei2013eeg,lin2006distributed,Sarvas1987}, and other inherent limitations. These factors hinder the accurate capture of neural activity in speech-related brain regions, thereby reducing the quality of information available for effective decoding. Recent advances in deep learning have shown promise in modeling complex neural signals \cite{wang2026hierarchical,xu2024convconcatnet,accou2023decoding}. However, most existing models focus on intra-subject decoding and face several limitations, as illustrated on the left side of Fig. \ref{definition}(a). Substantial cross-subject variability in neural responses during speech processing \cite{myers2024individual,giovannone2021individual,huth2016natural} makes it difficult to train a decoding model using neural signals across subjects. Although some models introduce subject layers \cite{li2024visual,defossez2023decoding} to account for this variability, these approaches require simultaneous access to data from all subjects, as shown on the right side of Fig. \ref{definition}(a). This necessitates retraining the model when new subjects are added, resulting in high computational costs. Moreover, the latent representations of subject layers exhibit low consistency across subjects, complicating training with a unified encoder and limiting the model's overall performance, as shown in Fig. \ref{pesa}(a).

When identical stimuli are presented to different subjects, the shared components of their responses are likely to reflect task-related information \cite{hasson2004intersubject,dmochowski2012correlated,nastase2019measuring}, which is the most informative for cross-subject decoding. Models that more effectively extract these subject-consistent representations are therefore expected to generalize better to unseen subjects. However, existing methods typically lack explicit constraints to enforce cross-subject consistency, limiting both decoding performance and generalizability.

Motivated by these observations, we propose an innovative cross-subject perceived speech decoding (CPSD) framework. The motivation behind this framework is further exploring subject-consistent information. We introduce a novel method to remap data from different subjects into a standardized reference space, thereby enhancing subject consistency. This approach not only improves the extraction of task-related information but also enhances generalizability. To further adapt the model to individual subjects, we incorporate a personal specialization stage, which initializes the subject layer using the consistent components learned during pre-training and subsequently aligns the target subject's neural data with the corresponding speech representations.

To address these issues, the key contributions of our work are as follows:
\begin{itemize}
    \item A \textbf{C}ross-Subject \textbf{P}erceived \textbf{S}peech \textbf{D}ecoding (\textbf{CPSD}) framework is proposed. To the best of our knowledge, this is the first cross-subject decoding framework specifically designed for perceived speech decoding. The CPSD framework consists of two training stages: source model pre-training and personal specialization.
    \item In this framework, a \textbf{P}ositional \textbf{E}ncoding-based \textbf{S}patial \textbf{A}ttention (\textbf{PESA}) module and a subject layer initialization method are introduced to explicitly extract subject-consistent information, thereby enabling effective cross-subject decoding. 
    \item The proposed framework was evaluated on three MEG/EEG datasets encompassing different modalities and languages. The results demonstrate consistent performance improvements over baseline methods across subjects. Ablation studies further confirm the effectiveness of the PESA module and the personal specialization stage.
    \item We conducted extensive analyses, including neural representation consistency evaluation, training time comparison, zero-shot decoding, hyperparameter search, comparison with multi-subject decoding, and PESA analysis. Collectively, these analyses demonstrate the effectiveness, efficiency, and robustness of the proposed framework.
\end{itemize}

\section{Related Work}
\subsection{Perceived Speech Decoding}
Existing studies include category classification from EEG \cite{simanova2010identifying}, achieving a decoding accuracy of 61\% on this binary classification task. However, this task is constrained by a strict experimental paradigm that differs significantly from the natural language used in daily life. In the context of continuous perceived speech decoding, fMRI has been employed to decode perceived speech for text generation \cite{lu2026brain,chen2024open,tang2023semantic}. Nevertheless, due to its low temporal resolution, fMRI struggles to fully capture dynamic information and tends to produce generated texts with a high word error rate. VLAAI \cite{accou2023decoding} reconstructs the speech envelope from EEG signals; however, this method focuses solely on low-level speech features, limiting its decoding performance. Additionally, previous studies have not fully leveraged the relationships between data from different subjects within the dataset. Brainmagic \cite{defossez2023decoding} utilizes wav2vec representations as decoding targets and employs a multi-subject training strategy, resulting in significant advancements in perceived speech decoding. However, this method requires simultaneous use of data from all subjects, which hinders its ability to generalize efficiently to new subjects. Furthermore, although Brainmagic employs subject layers to account for subject variability, the latent representations from these layers exhibit low consistency, increasing the difficulty of training with a unified encoder and restricting the model's performance, as shown in Fig. \ref{pesa}(a).

\begin{figure*}[t]
\centering
\includegraphics[width=18cm, height=7.87cm]{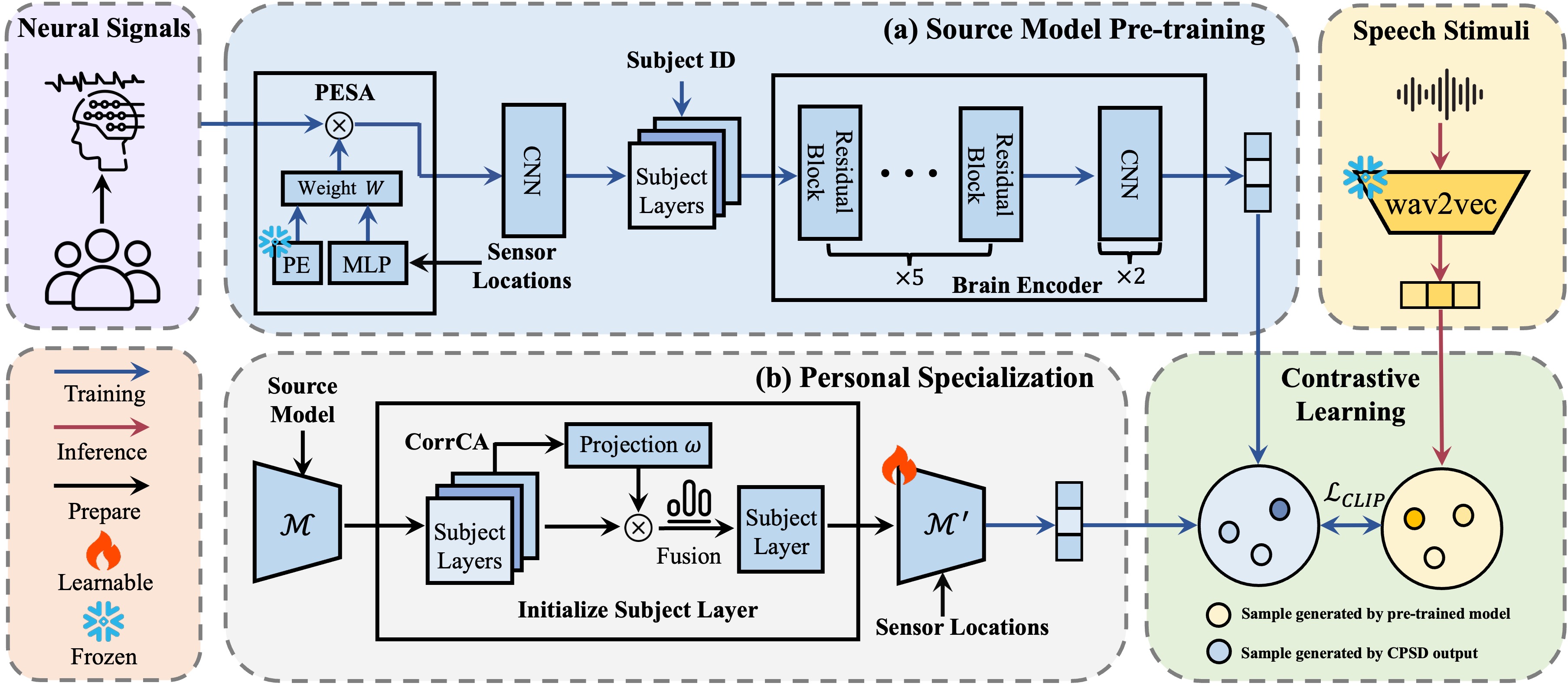}
\caption{Illustration of the CPSD framework. (a) During the source model pre-training stage, the model output is aligned with the wav2vec representation. (b) In the personal specialization stage, the subject layer for the target subject is first initialized, and then the representation from the MEG/EEG data of the target subject is realigned with the corresponding wav2vec representation.}
\label{CPSD}
\end{figure*}

\subsection{Cross-Subject Decoding}
Cross-subject decoding has been extensively studied and holds significant practical importance across various applications. One effective approach to addressing this challenge is acquiring transferable representations through self-supervised learning. TF-C \cite{zhang2022self} leverages the relationship between time and frequency to facilitate transfer between different datasets. Additionally, masked series modeling \cite{yang2023biot,dong2023simmtm} offers an alternative self-supervised learning approach by employing simple yet effective heuristic methods. Although it is widely acknowledged that self-supervised methods can learn generalizable representations to enhance performance on downstream tasks, these methods require the design of effective data augmentation techniques. However, this remains challenging for MEG/EEG speech decoding due to complex physical interpretations and low SNR. Domain adaptation methods, which learn through adversarial optimization between the encoder and a discriminator, aim to generate subject-invariant representations that can confuse the discriminator \cite{bethge2022domain,wang2024dmmr,ma2019reducing}, thereby facilitating transfer across subjects. However, acquiring source data can sometimes be more challenging than obtaining pre-trained models, creating a need to fine-tune the model solely on the target data. CL-CS \cite{hu2025cross} and CL-SSTER \cite{shen2024contrastive} utilize contrastive learning to maximize the similarity among different subjects exposed to the same stimulus, thereby acquiring MEG/EEG representations with higher subject consistency. Although this method is applicable to perceived speech, it lacks a design tailored to the corresponding decoding tasks. While MEG/EEG cross-subject decoding has been extensively studied, most related work has focused on emotion recognition, image decoding, and sleep staging. To the best of our knowledge, no framework has been specifically designed for cross-subject perceived speech decoding, a critical area for advancing research in speech processing. Additionally, current mainstream speech decoding models \cite{defossez2023decoding,accou2023decoding} often rely on specific architectures and therefore cannot directly adopt the aforementioned methods.

\section{Methodology}
\subsection{Problem Formulation}
Assume that $\{\mathcal{X}_{\mathcal{S}}^{i}\}_{i=1}^{N}$ and $\{\mathcal{X}_{\mathcal{S}^{'}}^{i}\}_{i=1}^{N}$ represent the source and target data, respectively, where $i$ indexes the MEG/EEG segments, and $\mathcal{S} \in \{1, 2, ..., I-1\}$, $\mathcal{S^{'}}=I$ denote the source and target subject indices. We iteratively select each subject in the dataset as the target subject. Both sets of neural data were evoked by the same stimuli, characterized by features $\{\mathcal{Y}^{j}\}_{j=1}^{N}$. The goal is to efficiently fine-tune the model $\mathcal{M}$, which has been pre-trained on the source data, to achieve high-performance match-mismatch classification on the target data. The task process is illustrated in Fig. \ref{definition}(b).

\subsection{Model Overview}

The model architecture of the CPSD framework is illustrated in Fig. \ref{CPSD}. The input data are first mapped to a standardized reference space using the PESA module, followed by a convolutional layer with a kernel size of 1. To fully leverage the source data, subject layers \cite{defossez2023decoding} are employed, and the latent representations are processed by the brain encoder to generate the final outputs. 

\begin{figure}[ht]
\centering
\includegraphics[width=8.8cm, height=2.15cm]{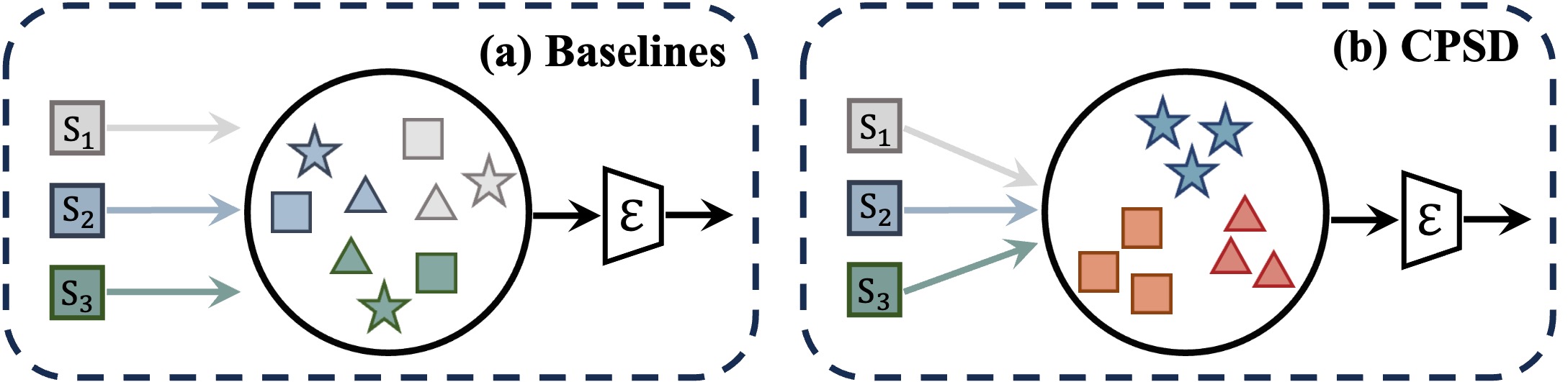}
\caption{Motivation for the PESA module. Different shapes within the large circle represent the MEG/EEG features of subjects under various stimuli.}
\label{pesa}
\end{figure}

\subsubsection{PESA}
To extract subject-consistent information more effectively and facilitate model training, we remap the MEG/EEG segments into a standardized reference space, as illustrated in Fig. \ref{pesa}(b). We select $D$ points in a high-dimensional space to serve as the reference. The positions of these points must satisfy the following properties: (1) each position has a unique embedding vector; (2) the design enables the model to attend to relative offsets rather than absolute positions; and (3) the embedding vectors effectively represent the positions. Positional encoding \cite{c:22} successfully meets these three properties and is commonly used to supplement positional information in transformer inputs. In our framework, we propose a novel application of positional encoding that utilizes embedding vectors to establish the reference space. The formula for positional encoding is shown in Eq. \eqref{pe}, where $r$ denotes the vectors, and $n$ and $i$ represent the indices of the vector and its dimension, respectively. We uniformly select $D$ embedding vectors, each with dimension $T$, with an interval of $\delta$ between adjacent vectors. In our experiments, $D$ is set to 270, consistent with \cite{defossez2023decoding}, and $\delta$ is 15.

\begin{equation}
r_{ni}= \label{pe}
\begin{cases}
    \text{sin}(\frac{n}{L^{i/D}})\text{, \quad if i is even,} \\
    \text{cos}(\frac{n}{L^{(i-1)/D}})\text{, if i is odd.}
\end{cases}
\end{equation}

Next, we need to obtain the weighting coefficients for the different channels. Sensor locations $X_{location}\in \mathbb{R}^{C\times 2}$ obtained from the MNE-Python function find\_layout \cite{gramfort2013meg} are used to represent the positional information of specific MEG/EEG systems. We then define a multilayer perceptron (MLP) to map these locations into a latent space with dimensions $C\times T$, as shown in Eq. \eqref{mlp}.

\begin{equation}
    Y = \text{MLP}(X_{location}), Y\in\mathbb{R}^{C\times T} \label{mlp}
\end{equation}

The MLP consists of three blocks, each comprising a linear layer, layer normalization \cite{ba2016layer}, and a GELU activation function \cite{hendrycks2016gaussian}. The input and output dimensions of the linear layers are set to $T$, except for the first layer, which must match the dimension of the input. Then, the similarity between the MLP output and the positional encoding is calculated to measure the contribution of the input data to different positions in the reference space. Subsequently, the softmax function is applied to obtain the final input weights, as shown in Eq. \eqref{W}, where $P=(r_{n\delta,i})_{n=1,...,D}^{i=1,...,T}$ denotes the stack of positional embedding vectors.

\begin{equation}
    W = \text{Softmax}(YP^{T}), W\in\mathbb{R}^{C\times D} \label{W}
\end{equation}

Lastly, the latent representation $H$ is computed from the input $X\in\mathbb{R}^{C\times T}$ using the formula in Eq. \eqref{final}.

\begin{equation}
    H = W^{T}X,H\in\mathbb{R}^{D\times T}\label{final}
\end{equation} 

\subsubsection{Brain Encoder}
Due to the superior decoding performance of Brainmagic \cite{defossez2023decoding}, we employ the same architecture for our encoder. The output of the subject layers is first processed through five residual blocks, each containing three convolutional layers. The first and second convolutional layers have input and output channels of 320, except for the first layer of the brain encoder. The dilation parameters are set to be $2^{2l\%5}$ and $2^{(2l+1)\%5}$, where $l$ is the residual block index, to increase the receptive field of these two convolutional layers. Both convolutional layers are followed by a GELU activation function and batch normalization \cite{ioffe2015batch}. The last convolutional layer in each block has an output channel size of 640, and a GLU activation function \cite{dauphin2017language} is used to reduce the dimension back to 320. The kernel size and stride parameters are set to 3 and 1, respectively, for all convolutions. Padding is applied to maintain the temporal dimension $T$. Finally, two convolutional layers are used, each followed by a GELU activation function. The first convolutional layer increases the input channel size from 320 to 640 with a kernel size of 1, and the subsequent $1\times 1$ convolutional layer adjusts the dimension to match that of the wav2vec representations.

\subsection{Training Pipeline}

The CPSD framework involves a two-stage training process: first, pre-training a source model to acquire prior knowledge by fully leveraging data from various subjects; second, personal specialization on the target data to adapt the model to the specific subject, as shown in Algorithm \ref{alg:alg1}:

\begin{algorithm}[H]
\caption{Cross-Subject Perceived Speech Decoding.}\label{alg:alg1}
\begin{algorithmic}
\STATE {\textbf{\textsc{Initialization}}}:
\STATE \hspace{0.2cm} 1: \hspace{0.2cm} Use a leave-one-subject-out approach, treating the remaining subjects in the dataset as source subjects.
\STATE \hspace{0.2cm} 2: \hspace{0.2cm} Get sensor locations using mne.find\_layout().
\STATE \hspace{0.2cm} 3: \hspace{0.2cm} Initialize the parameters of CPSD model $\mathcal{M}$.
\STATE 
\STATE {\textbf{\textsc{SOURCE MODEL PRE-TRAINING}}}:
\STATE \textbf{Input:} Initialized model $\mathcal{M}$, MEG/EEG segments, sensor locations $L$, subject ID $sid$, wav2vec representations $W$ and number of epochs $num_{1}$.
\STATE \textbf{Output:} Pre-trained source model $\mathcal{M}_{smp}$.
\STATE \hspace{0.2cm} 1: \hspace{0.2cm} \textbf{for} $i=1$ to $num_{1}$ \textbf{do}
\STATE \hspace{0.2cm} 2: \hspace{0.6cm} $H \gets \mathcal{M}$(MEG/EEG, $L$, $sid$)
\STATE \hspace{0.2cm} 3: \hspace{0.6cm} $Loss \gets \mathcal{L}_{CLIP}$($H$, $W$)
\STATE \hspace{0.2cm} 4: \hspace{0.6cm} Optimize the parameters of $\mathcal{M}$.
\STATE \hspace{0.2cm} 5: \hspace{0.2cm} \textbf{end for}
\STATE \hspace{0.2cm} 6: \hspace{0.2cm} \textbf{return} $\mathcal{M}_{smp}$
\STATE 
\STATE {\textbf{\textsc{PERSONAL SPECIALIZATION}}}:
\STATE \textbf{Input:} Source model $\mathcal{M}_{smp}$, MEG/EEG segments, sensor locations $L$, wav2vec representations $W$ and number of epochs $num_{2}$.
\STATE \textbf{Output:} Personal specialized model $\mathcal{M}_{ps}$ on the target subject.
\STATE \hspace{0.2cm} 1: \hspace{0.2cm} Initialize subject layer for target subject by Eq. \eqref{corrca} and \eqref{average}, then get model $\mathcal{M}_{smp}^{'}$.
\STATE \hspace{0.2cm} 2: \hspace{0.2cm} \textbf{for} $i=1$ to $num_{2}$ \textbf{do}
\STATE \hspace{0.2cm} 3: \hspace{0.6cm} $\mathbf{H} \gets \mathcal{M}_{smp}^{'}$(MEG/EEG, $L$)
\STATE \hspace{0.2cm} 4: \hspace{0.6cm} $Loss \gets \mathcal{L}_{CLIP}$($H$, $W$)
\STATE \hspace{0.2cm} 5: \hspace{0.6cm} Optimize the parameters of $\mathcal{M}_{smp}^{'}$.
\STATE \hspace{0.2cm} 6: \hspace{0.2cm} \textbf{end for}
\STATE \hspace{0.2cm} 7: \hspace{0.2cm} \textbf{return} $\mathcal{M}_{ps}$
\STATE
\STATE {\textbf{\textsc{EVALUATION}:}}
\STATE \textbf{Input:} MEG/EEG segments, sensor locations $L$ and wav2vec representations $W$.
\STATE \textbf{Output:} Top-k and rank accuracy.
\STATE \hspace{0.2cm} 1: \hspace{0.2cm} $\mathbf{H} \gets  \mathcal{M}_{ps}$(MEG/EEG,$L$)
\STATE \hspace{0.2cm} 2: \hspace{0.2cm} \textbf{return} Topk($H$, $W$), Rank($H$, $W$)
\end{algorithmic}
\label{alg1}
\end{algorithm}

\begin{table*}[h]
\caption{A brief description of the neural datasets related to speech perception. The test samples indicate the size of the retrieval set for each sample.}
\renewcommand{\arraystretch}{1.75}
\centering
\begin{tabular}{lcccccccccc}
\toprule
{Dataset} & {Language} & {Modality} & {Sensors} & {Sampling rate} & {Subjects} & {Folds for CV} & {Duration} & {Test samples}\\
\midrule
Armeni 2022 & English & MEG & 275 & 1200 & 3 & 3 & 30.0 h & 1024\\
PKUEEG 2025 & Chinese & EEG & 64 & 1000 & 25 & 25 & 68.1 h & 128\\
Broderick 2018 & English & EEG & 128 & 512 & 19 & 19 & 19.2 h & 128\\
\bottomrule
\end{tabular}
\label{dataset}
\end{table*}

\subsubsection{Source Model Pre-Training}
The model $\mathcal{M}$ in the CPSD framework is initially pre-trained on the source data, which includes $I-1$ subjects. Model $\mathcal{M}$ requires three distinct inputs: the source data, sensor locations, and subject IDs. The source data and sensor locations are first fed into the PESA module, followed by a $1\times 1$ convolutional layer. Subsequently, subject layers process the latent representations along with the subject IDs. Data from different subjects at the same index are randomly selected to facilitate multi-subject training. Finally, a brain encoder produces the final output, which is aligned with the wav2vec representation extracted from the corresponding stimuli.

\subsubsection{Personal Specialization} 
To fine-tune the pre-trained model $\mathcal{M}$ for the target subject, we first initialize a subject layer with parameters tailored to that subject. We continue to leverage subject consistency to address this challenge; the consistent components of the subject layers are expected to capture the shared information across subjects in our decoding task. Therefore, we employ CorrCA \cite{parra2018correlated,dmochowski2012correlated} to initialize the target subject layer based on the subject layers of the source subjects. The computational steps of the CorrCA algorithm are detailed in Eq. \eqref{corrca}:

\begin{equation}
    \label{corrca}
    \hat{w}=\mathop{\arg\max}\limits_{w} \frac{w^{T}X_{1}X_{2}^{T}w}{\Vert X_{1}^{T}w \Vert \Vert X_{2}^{T}w \Vert} 
\end{equation}

\noindent where $X_{1}$, $X_{2}$, and $w$ denote two input data matrices and the weight matrix, respectively, all belonging to $\mathbb{R}^{D\times D}$. The goal of CorrCA is to optimize the weight matrix to maximize the correlation between subjects. Next, a new model $\mathcal{M}^{'}$ is initialized by replacing the subject layers with the average of the consistent components across subjects, as shown in Eq. \eqref{average}: 

\begin{equation}
    \label{average}
    X_{I} = \frac{1}{I-1}\sum_{i=1}^{I-1}X_i^{T}\hat{w}
\end{equation}

\noindent where $I$ is the index of the target subject, and $\{1,...,I-1\}$ are the indices of the source subjects. Finally, we realign the model outputs from the target subject with the corresponding wav2vec representations to specifically adapt the model for that subject.

\begin{table*}[h]
\caption{Perceived speech decoding results across various datasets. We averaged the decoding results across different subjects and evaluated the robustness of performance improvements using statistical tests.}
\renewcommand{\arraystretch}{1.75}
\centering
\resizebox{2\columnwidth}{!}{
\begin{tabular}{l|ll|ll|ll}
\toprule
\multicolumn{1}{c}{} & \multicolumn{2}{c}{Armeni 2022} & \multicolumn{2}{c}{PKUEEG 2025} & \multicolumn{2}{c}{Broderick 2018} \\
\cmidrule(rl){2-3} \cmidrule(rl){4-5} \cmidrule(rl){6-7}
Methods & Top-10(\%) & Rankacc(\%) & Top-10(\%) & Rankacc(\%) & Top-10(\%) & Rankacc(\%) \\
\midrule
Random & 1.3 $\pm$ 0.4 &  50.6 $\pm$ 0.5 & 7.3 $\pm$ 1.2 &  49.4 $\pm$ 2.2 & 7.6 $\pm$ 1.6 &  50.8 $\pm$ 0.3 \\
\midrule
ATM-S \cite{li2024visual} & 4.2 $\pm$ 1.4 & 68.8 $\pm$ 4.8 & 9.2 $\pm$ 3.2 & 52.7 $\pm$ 2.1 & 9.0 $\pm$ 2.8 & 51.9 $\pm$ 3.3 \\
iTransformer \cite{liuitransformer} & 15.0 $\pm$ 12.3 &  76.6 $\pm$ 16.4 & 11.5 $\pm$ 3.5 &  56.3 $\pm$ 3.1 & 9.3 $\pm$ 2.9 &   52.1 $\pm$ 3.3 \\
VLAAI \cite{accou2023decoding} & 14.7 $\pm$ 13.2 &  75.0 $\pm$ 19.4 & 14.6 $\pm$ 12.4 &  55.9 $\pm$ 11.3 & 10.5 $\pm$ 4.9 &  55.5 $\pm$ 6.5 \\
Brainmagic \cite{defossez2023decoding} & 17.4 $\pm$ 13.8 &   83.1 $\pm$ 7.7 & 20.8 $\pm$ 8.1 &   65.7 $\pm$ 7.0 & 17.7 $\pm$ 7.1 &   63.0 $\pm$ 5.5 \\
\midrule
BIOT \cite{yang2023biot} & 23.8 $\pm$ 16.5 &  86.8 $\pm$ 6.8 & 22.9 $\pm$ 9.7 &  67.0 $\pm$ 8.0 & 20.0 $\pm$ 8.6 &  63.9 $\pm$ 7.3 \\
DAPE \cite{bethge2022domain}  & 51.5 $\pm$ 13.6 & 95.4 $\pm$ 1.9 & 25.9 $\pm$ 12.7 &   69.1 $\pm$ 8.6 & 22.5 $\pm$ 8.5 &   66.5 $\pm$ 6.4 \\
CL-CS \cite{hu2025cross}  & 54.5 $\pm$ 9.7 &  96.1 $\pm$ 1.5 & 27.6 $\pm$ 10.0 &  70.8 $\pm$ 7.5 & 24.1 $\pm$ 9.6 & 68.9 $\pm$ 6.5 \\
\textbf{CPSD (Ours)} & \textbf{61.3} $\pm$ 12.4 &  \textbf{96.9} $\pm$ 1.5 & \textbf{43.0} $\pm$ 13.6 &  \textbf{80.0} $\pm$ 6.7 & \textbf{39.9} $\pm$ 15.3 & \textbf{77.7} $\pm$ 8.0 \\
\bottomrule
\end{tabular}
}
\label{main}
\end{table*}

\subsection{Loss Function}
In our experiments, we employed the CLIP loss \cite{radford2021learning} during both the source model pre-training stage and the personal specialization stage. This loss function is widely used in various applications \cite{benchetrit2024brain,li2023blip,mu2022slip} and has demonstrated superior performance in alignment tasks. It aims to align neural representations with corresponding speech features by maximizing the similarity between positive pairs while minimizing the similarity between negative pairs. The loss function is calculated as shown in Eq. \eqref{clip}:

\begin{equation}
\mathcal{L}_{CLIP} = -\frac{1}{N} \sum_{i=1}^{N} \log \left( \frac{\exp \left( \text{sim}(z_i, y_i) \right)}{\sum_{j=1}^{N} \exp \left( \text{sim}(z_i, y_j) \right)} \right) \label{clip}
\end{equation}

\noindent where $\mathcal{L}_{CLIP}$ represents the loss function, $N$ represents the batch size, $z_{i}$ represents the model outputs, $y_{i}$ represents the wav2vec representations extracted from the speech segments, and sim($x$, $y$) computes the similarity between $x$ and $y$.

\section{Experiments}
\subsection{Datasets}
Our framework was evaluated on three neural datasets related to speech perception: Armeni 2022 \cite{armeni202210}, PKUEEG 2025, and Broderick 2018 \cite{broderick2018electrophysiological}, as summarized in Table \ref{dataset}. \textbf{Armeni 2022}: This public dataset includes data from three subjects who listened to English stories for a total of 30 hours. MEG data were recorded using a 275-channel axial gradiometer CTF system with a sampling rate of 1200 Hz. \textbf{PKUEEG 2025}: This dataset involves 25 subjects listening to the Chinese story \textit{Romance of the Three Kingdoms}. Sixty-four-channel EEG data were recorded using the NeuroScan system at 1000 Hz. \textbf{Broderick 2018}: This public dataset contains EEG data from 19 English-speaking subjects listening to extracts from \textit{The Old Man and the Sea}. A BioSemi ActiveTwo system with 128 channels was used, and the sampling rate was 512 Hz.

\subsection{Data Preprocessing}
For the MEG/EEG data, a notch filter was initially applied to eliminate line noise. The data were then resampled to 100 Hz to extract relevant information from the low-frequency band for our decoding task \cite{hamalainen1993magnetoencephalography}. Independent Component Analysis (ICA) was performed on each trial to remove eye-blink artifacts \cite{jung1998independent}. The resampled data were segmented into 3-second intervals with a 1.5-second overlap between consecutive segments. Subsequently, the segments were normalized using RobustScaler, and values below -20 or above 20 were clamped to mitigate the influence of outliers \cite{defossez2023decoding}. For our match-mismatch classification task, the wav2vec representation was selected as the feature representing the speech stimulus \cite{defossez2023decoding}, extracted from the final output of wav2vec2-large-xlsr-53 \cite{baevski2020wav2vec}.

\subsection{Baseline Methods} 
Baseline methods were selected from two main categories: intra-subject decoding and cross-subject decoding. Since cross-subject perceived speech decoding has not been addressed in previous work, models and frameworks from related fields were also considered to provide a comprehensive comparison with our framework. For intra-subject decoding, ATM-S \cite{li2024visual}, iTransformer \cite{liuitransformer}, VLAAI \cite{accou2023decoding}, and Brainmagic \cite{defossez2023decoding} were chosen, with the best-performing model selected as the encoder. To compare different methods in cross-subject decoding, BIOT \cite{yang2023biot}, DAPE \cite{bethge2022domain}, and CL-CS \cite{hu2025cross} were selected as baseline methods. For all baselines, training stage hyperparameters were set according to the values recommended in their original papers to ensure optimal performance. Under cross-subject conditions, BIOT, DAPE, and CL-CS use the same training stages as CPSD.

\begin{table*}[h]
\caption{Ablation study of the CPSD framework across different datasets.\\(PS stands for personal specialization.)}
\renewcommand{\arraystretch}{1.75}
\centering
\resizebox{2\columnwidth}{!}{
\begin{tabular}{l|ll|ll|ll}
\toprule
\multicolumn{1}{c}{} & \multicolumn{2}{c}{Armeni 2022} & \multicolumn{2}{c}{PKUEEG 2025} & \multicolumn{2}{c}{Broderick 2018}\\
\cmidrule(rl){2-3} \cmidrule(rl){4-5} \cmidrule(rl){6-7}
\textbf{}& Top-10(\%) & Rankacc(\%) & Top-10(\%) & Rankacc(\%) & Top-10(\%) & Rankacc(\%)\\
\midrule
Base & 17.4 $\pm$ 13.8 &  83.1 $\pm$ 7.7 & 20.8 $\pm$ 8.1 &  65.7 $\pm$ 7.0 & 17.7 $\pm$ 7.1 & 63.0 $\pm$ 5.5 \\
Base+PESA &  25.7 $\pm$ 25.5 &  84.7 $\pm$ 9.8 & 27.7 $\pm$ 8.5 &  70.7 $\pm$ 5.7 & 18.1 $\pm$ 8.3 &  62.6 $\pm$ 6.9 \\
Base+PS &  50.3 $\pm$ 12.8 &  95.1 $\pm$ 2.2 &  29.2 $\pm$ 10.9 &  72.2 $\pm$ 6.3 &  35.3 $\pm$ 11.8 &  75.2 $\pm$ 7.0 \\
\textbf{Base+PESA+PS} & \textbf{61.3} $\pm$ 12.4 &  \textbf{96.9} $\pm$ 1.5 & \textbf{43.0} $\pm$ 13.6 &  \textbf{80.0} $\pm$ 6.7 & \textbf{39.9} $\pm$ 15.3 & \textbf{77.7} $\pm$ 8.0 \\
\bottomrule
\end{tabular}
}
\label{ablation}
\end{table*}

\subsection{Implementation Details}
Data from both source and target subjects are partitioned using the same method. To avoid data leakage \cite{xu2026impacts,li2020perils}, the dataset is divided into 70\%, 10\%, and 20\% splits based on the order of trials. The stimuli for trials in the training, validation, and test sets do not overlap. The mini-batch size is set to 128 during both the source model pre-training stage and the personal specialization stage. The Adam optimizer \cite{kingma2014adam} is used to minimize the loss, with a learning rate of $3\times 10^{-4}$ in both stages. To prevent overfitting on the training set, early stopping is applied if the model's performance does not improve for 10 consecutive epochs on the validation set. The model with the best validation performance is then selected for personal specialization or evaluation. The chance level for these experiments is provided in the main results Table \ref{main}. We first randomly generate embeddings with the same shape as the wav2vec vectors and evaluate the results on the test set. All experiments are conducted on a single NVIDIA RTX 3090 GPU.

\subsection{Evaluation Metrics}
In this study, various evaluation metrics were employed to comprehensively assess the model's performance. Top-10 accuracy was used as the evaluation metric for the match-mismatch classification tasks, measuring whether the target segment appeared among the model's ten most likely predictions, as shown in Eq. \eqref{topk}:

\begin{equation}
    \label{topk}
    Top\text{-}10 = \frac{1}{N}\sum_{i=1}^{N}\mathbb{I}(l_{i} \in \hat{l}_{i,1},...,\hat{l}_{i,10})
\end{equation}

\noindent where $N$ represents the number of test samples, $\mathbb{I}$ denotes the indicator function, $l_{i}$ is the true label of the $i$-th sample, and $\{\hat{l}_{i,1},...,\hat{l}_{i,10}\}$ are the labels sorted in descending order according to their predicted probabilities. Additionally, rank accuracy was used to evaluate the predictive performance on target samples in the test set. The formula for calculating rank accuracy is shown in Eq. \eqref{rankacc}:

\begin{equation}
Rankacc=1-\frac{<rank> - 1}{<segments> - 1} \label{rankacc}
\end{equation}

\noindent where $<rank>$ represents the position of the corresponding target sample among the test samples, and $<segments>$ denotes the total number of test samples. These two evaluation metrics are commonly used in speech decoding tasks \cite{defossez2023decoding,pereira2018toward}, facilitating comparison with prior work.

\subsection{Experimental Results}
\subsubsection{Main Results}
The results of perceived speech decoding are presented in Table \ref{main}. We first compare the performance of different encoders, including ATM-S, iTransformer, VLAAI, and Brainmagic. Brainmagic demonstrates superior performance compared to the other model architectures when decoding data exclusively from the target subject. These results validate our choice of the Brain Encoder. To compare different cross-subject decoding methods, we consider BIOT, DAPE, CL-CS, and CPSD. Our framework surpasses these baseline methods across all three datasets, achieving Top-10 accuracies of 61.3\%, 43.0\%, and 39.9\%, respectively. A pairwise t-test was also conducted, indicating that the improvements are statistically significant for each dataset ($p < 0.001$), demonstrating the reliability of our results.

\subsubsection{Ablation Study}
The results of the ablation study are presented in Table \ref{ablation}. To demonstrate the effectiveness of the PESA module and the personal specialization stage, four settings are compared, defined as follows:
\begin{itemize}
	\item Base: The model was trained and evaluated exclusively on the target data (intra-subject decoding) without utilizing the PESA module.
	\item Base+PESA: The CPSD framework was trained and evaluated exclusively on the target data.
	\item Base+PS: CPSD framework was implemented without the PESA module but included two training stages: source model pre-training and personal specialization.
	\item Base+PESA+PS: We utilized the complete version of the CPSD framework.
\end{itemize}

The results presented in the first two rows indicate that implementing the PESA module enhances the model's performance in intra-subject decoding for the target subject, thereby validating the initial assumption that PESA optimizes the training process. The two-stage training approach demonstrates significant improvements across three datasets, showing that our method effectively leverages prior knowledge from source subjects' data to facilitate generalization to the target subject. Ultimately, training with the full CPSD framework yields the highest decoding accuracies, confirming that the PESA module improves the model's generalizability. Additionally, pairwise t-tests reveal that all these improvements are statistically significant ($p < 0.001$).

\begin{table*}[h]
\caption{Evaluation of neural representation consistency.}
\renewcommand{\arraystretch}{1.75}
\centering
\resizebox{2\columnwidth}{!}{
\begin{tabular}{l|cc|cc|cc}
\toprule
\multicolumn{1}{c}{} & \multicolumn{2}{c}{Armeni 2022} & \multicolumn{2}{c}{PKUEEG 2025} & \multicolumn{2}{c}{Broderick 2018}\\
\cmidrule(rl){2-3} \cmidrule(rl){4-5} \cmidrule(rl){6-7}
Methods & Latent & Final & Latent & Final & Latent & Final\\
\midrule
Brainmagic & 0.011 $\pm$ 0.001 &  0.230 $\pm$ 0.012 & \textbf{0.003} $\pm$ 0.001 & 0.093 $\pm$ 0.022 & 0.001 $\pm$ 0.001 & 0.157 $\pm$ 0.009\\
\textbf{CPSD(Ours)} & \textbf{0.016} $\pm$ 0.002 & \textbf{0.274} $\pm$ 0.012 & 0.002 $\pm$ 0.001 & \textbf{0.129} $\pm$ 0.010 & \textbf{0.002} $\pm$ 0.001 & \textbf{0.227} $\pm$ 0.010\\
\bottomrule
\end{tabular}
}
\label{iscc}
\end{table*}

\subsubsection{Neural Representation Consistency}
The subject-consistent information extraction capabilities of pre-trained source models were evaluated using the inter-subject correlation (ISC) metric, calculated via the CorrCA algorithm. The results are presented in Table \ref{iscc}. Two experimental settings were examined: the latent representation from the subject layer and the final output. Compared to Brainmagic, the CPSD framework demonstrated improvements in nearly all settings, with results that were statistically significant ($p<0.001$). 

\begin{figure}[ht]
\centering
\includegraphics[width=8.8cm, height=3.52cm]{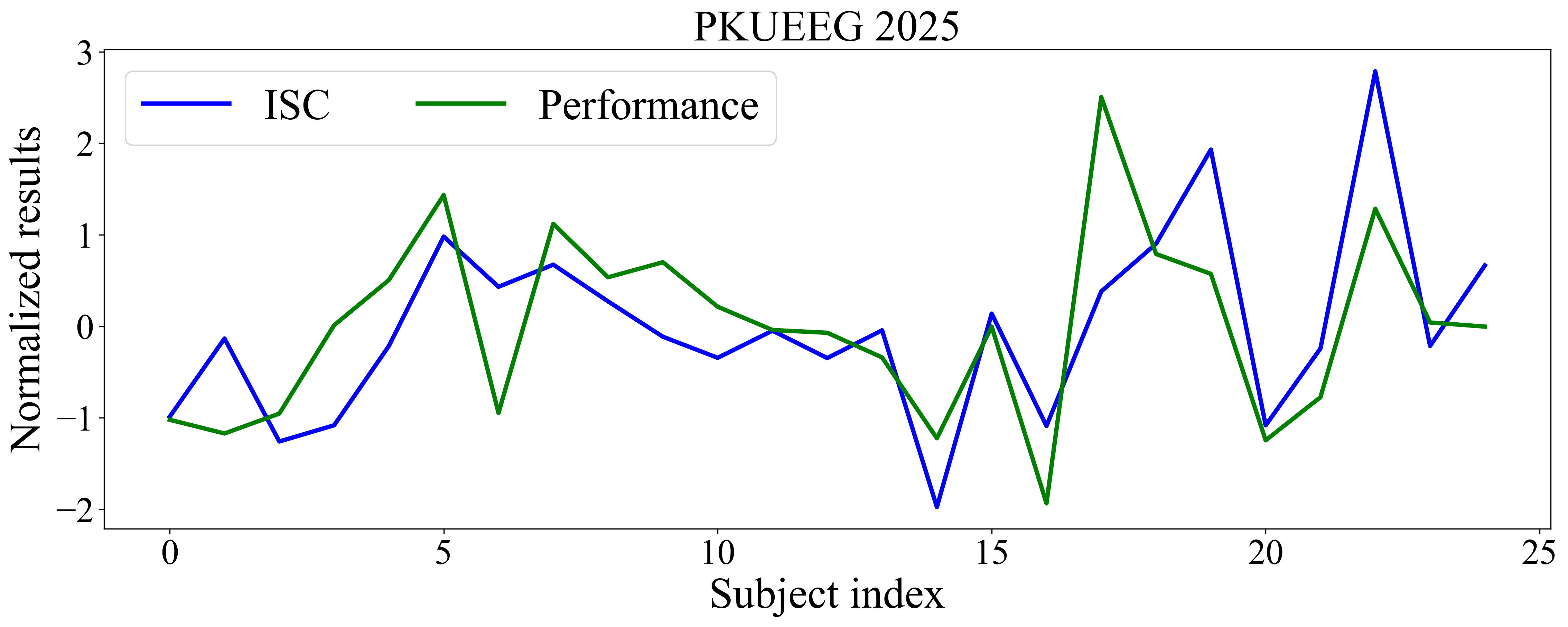}
\caption{The correlation between model performance and neural representation consistency ($corr=0.66$, $p<0.001$).}
\label{isc}
\end{figure}

Although Brainmagic exhibited higher consistency in the subject layer output of the PKUEEG 2025 dataset, the p-value for this result exceeded 0.05. Additionally, we provide the correlation between normalized Top-10 accuracy and normalized ISC on the PKUEEG 2025 dataset, as shown in Fig. \ref{isc}.

The results indicate that the consistency of neural representations following pre-training is positively correlated with decoding performance, thereby supporting our hypothesis that incorporating subject-consistent information enhances decoding accuracy. This conclusion is further supported by the Broderick 2018 dataset, which demonstrates a correlation coefficient of 0.77.

\subsubsection{Training Time Comparison}
\begin{figure}[ht]
\centering
\includegraphics[width=8.8cm, height=4.46cm]{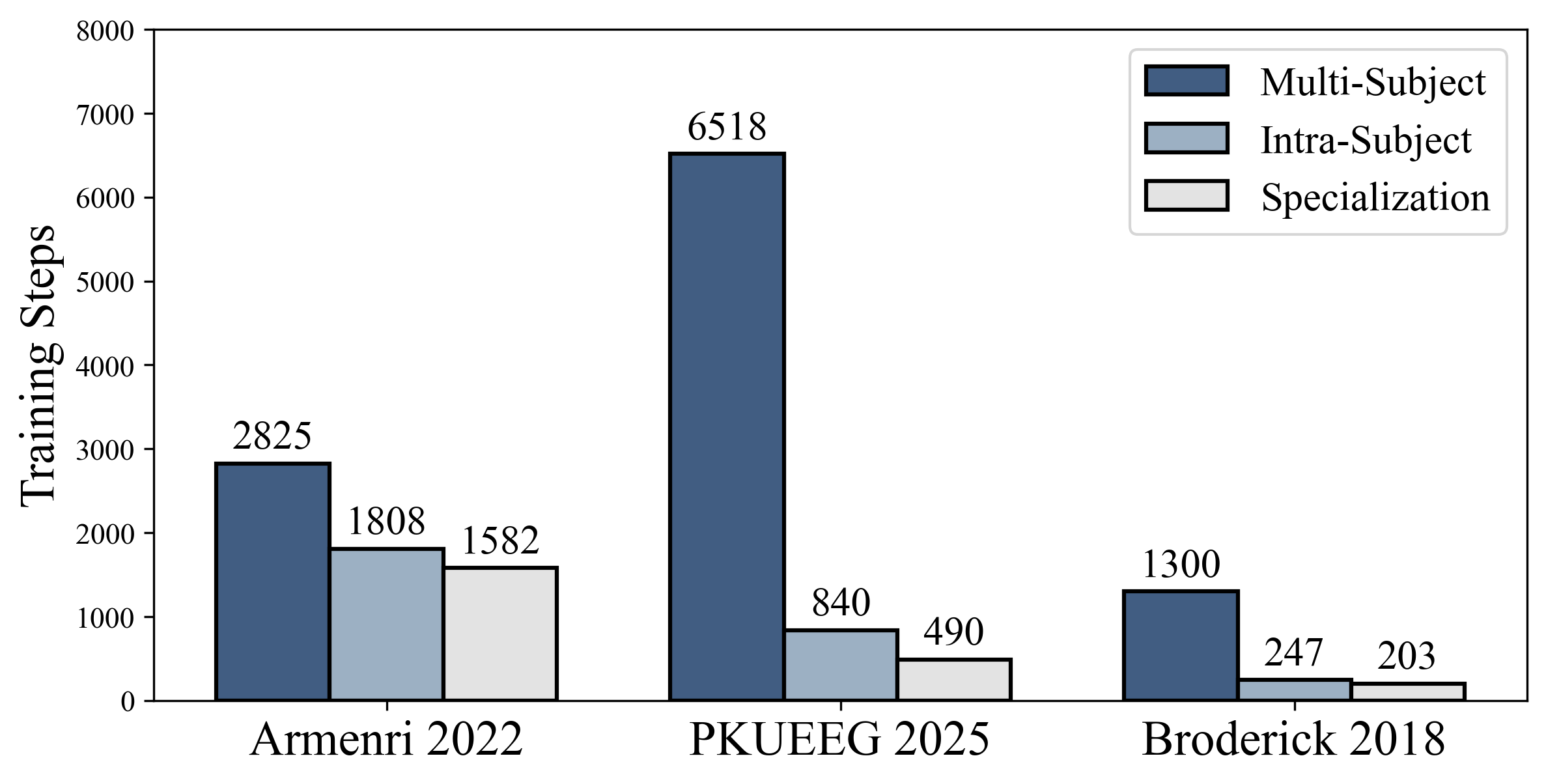}
\caption{Training times under different settings.}
\label{time}
\end{figure}

To demonstrate the generalization efficiency of CPSD for the target subject, we analyze its training costs under different settings, including multi-subject training, intra-subject training, and the personal specialization stage. The results are presented in Fig. \ref{time}. We use the number of training steps as the evaluation metric due to interference from other concurrently running processes. Our framework exhibits lower training costs compared to the other settings. This reduction in computational demand is more pronounced in datasets with a larger number of subjects. Specifically, the difference in training steps for the Armeni 2022 dataset is relatively minor, likely due to the small number of subjects in this dataset. In contrast, the specialization stage requires only 7.5\% and 15.6\% of the training steps needed for multi-subject training on the PKUEEG 2025 and Broderick 2018 datasets, respectively.

\subsubsection{Zero-Shot Decoding}
\begin{figure}[ht]
\centering
\includegraphics[width=8.8cm, height=4.37cm]{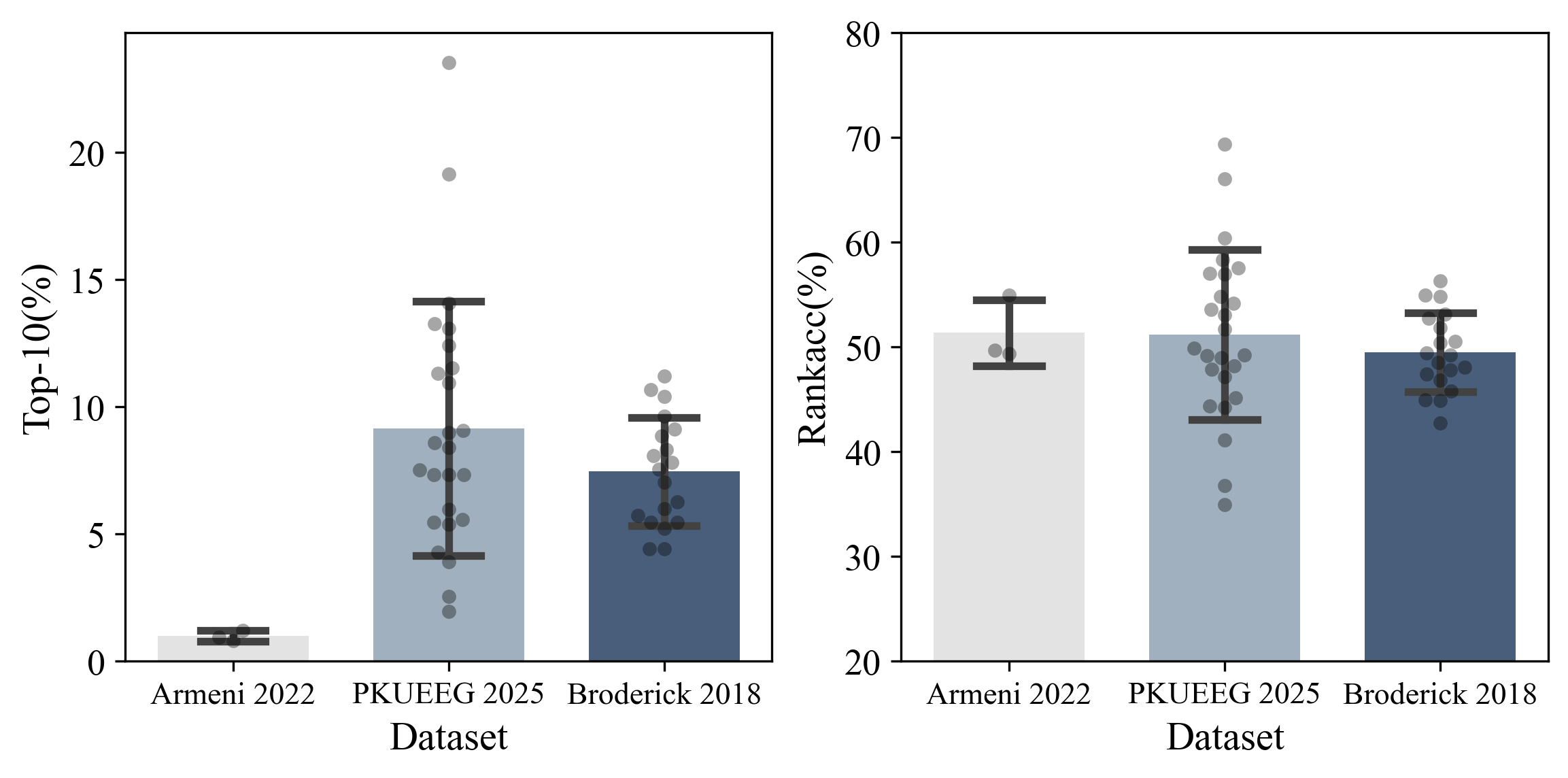}
\caption{Zero-shot decoding performance across various perceived speech datasets.}
\label{zeroshot}
\end{figure}

To further evaluate the generalizability of our CPSD framework, we present perceived speech decoding results under a zero-shot setting. In this scenario, we use only the source model pre-training stage and directly evaluate the framework's performance on the test set without any additional specialization. This approach is particularly useful when training data from the target subject are limited \cite{wang2024dmmr} and presents a challenging task due to substantial cross-subject variability, which has been underexplored in perceived speech decoding. As shown in Fig. \ref{zeroshot}, even without personal specialization, our model achieves decoding results significantly above chance level in 44.7\% (21/47) of subjects. The chance level of decoding is presented in Table \ref{main}. The proportion of subjects whose decoding accuracy exceeds the chance level is 33.3\% (1/3), 48.0\% (12/25), and 42.1\% (8/19), respectively. These results highlight the importance of the number of source subjects and the significance of the personal specialization stage for the model. This demonstrates that our framework effectively captures consistent representations from source subjects that generalize to the target subject, underscoring the robustness of our methods.

\subsubsection{Hyperparameter Search}
\begin{figure}[ht]
\centering
\includegraphics[width=8.8cm, height=3.65cm]{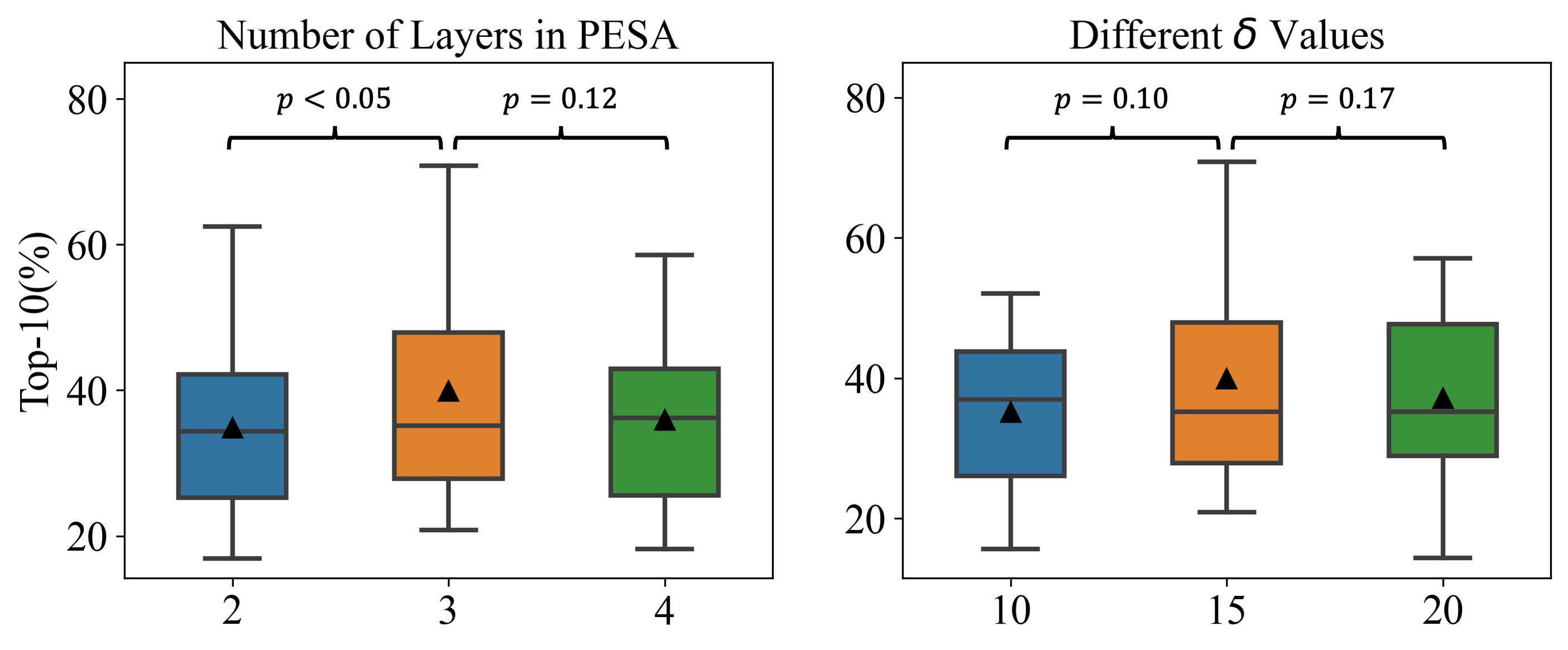}
\caption{Decoding results using various hyperparameters on the Broderick 2018 dataset. The triangles in the box plot indicate the mean values.}
\label{hyper}
\end{figure}

We also conducted hyperparameter searches to evaluate the reliability of our results. Our framework introduces two hyperparameters: $\alpha$, which denotes the number of MLP layers in PESA, and $\delta$, which represents the interval between adjacent position embeddings. Due to the high computational cost, we evaluated our framework only on the Broderick 2018 dataset. Fig. \ref{hyper} shows that the CPSD framework is not sensitive to hyperparameter selection, and we chose the hyperparameters that yielded the highest Top-10 accuracy. The experimental results in Fig. \ref{hyper} demonstrate that our framework maintains high performance under varying conditions, thereby confirming the robustness of our findings.

\begin{table*}[h]
\caption{Comparison with Multi-Subject Settings.\\(MD stands for Multi-Subject Decoding.)}
\renewcommand{\arraystretch}{1.5}
\centering
\resizebox{2\columnwidth}{!}{
\begin{tabular}{l|ll|ll|ll}
\toprule
\multicolumn{1}{c}{} & \multicolumn{2}{c}{Armeni 2022} & \multicolumn{2}{c}{PKUEEG 2025} & \multicolumn{2}{c}{Broderick 2018}\\
\cmidrule(rl){2-3} \cmidrule(rl){4-5} \cmidrule(rl){6-7}
Methods & Top-10(\%) & Rankacc(\%) & Top-10(\%) & Rankacc(\%) & Top-10(\%) & Rankacc(\%) \\
\midrule
Brainmagic& 17.4 $\pm$ 13.8 &   83.1 $\pm$ 7.7 & 20.8 $\pm$ 8.1 &   65.7 $\pm$ 7.0 & 17.7 $\pm$ 7.1 &   63.0 $\pm$ 5.5 \\
Brainmagic-MD& 52.7 $\pm$ 15.5 & 95.6 $\pm$ 2.1 & 39.1 $\pm$ 10.3 & 78.8 $\pm$ 5.3 & 39.2 $\pm$ 12.4 & 77.4 $\pm$ 7.1 \\
\midrule
CPSD & 61.3 $\pm$ 12.4 &  96.9 $\pm$ 1.5 & 43.0 $\pm$ 13.6 &  80.0 $\pm$ 6.7 & 39.9 $\pm$ 15.3 & 77.7 $\pm$ 8.0 \\
\textbf{CPSD-MD} & \textbf{63.0} $\pm$ 11.8 &  \textbf{97.1} $\pm$ 1.3 & \textbf{45.5} $\pm$ 11.5 &  \textbf{81.5} $\pm$ 5.5 & \textbf{46.0} $\pm$ 14.5 & \textbf{81.0} $\pm$ 6.6 \\
\bottomrule
\end{tabular}
}
\label{multi}
\end{table*}

\subsubsection{Comparison with Multi-Subject Settings}
To highlight the advancements of the PESA module, we evaluated the CPSD framework in a multi-subject setting, which has demonstrated effectiveness in decoding perceived speech \cite{defossez2023decoding}. We selected Brainmagic and CPSD as our comparison targets and trained models on all subjects in the dataset simultaneously. By fully leveraging contrastive learning, both models showed improved decoding performance in multi-subject settings, as shown in Table \ref{multi}. However, the performance degradation of the CPSD framework was significantly smaller than that of Brainmagic. Compared to the decoding results obtained from multi-subject training, our cross-subject results showed only a performance decrease of 1.7\%, 2.5\%, and 6.1\%, respectively, highlighting the advancement of our source model pre-training stage. When both CPSD and Brainmagic were applied to multi-subject decoding, the Top-10 accuracy improved by 10.3\%, 6.4\%, and 6.8\% across the three datasets, further demonstrating the advantages of our model architecture design.


\subsubsection{PESA Analysis}
\begin{figure}[ht]
\centering
\includegraphics[width=8.8cm, height=3.93cm]{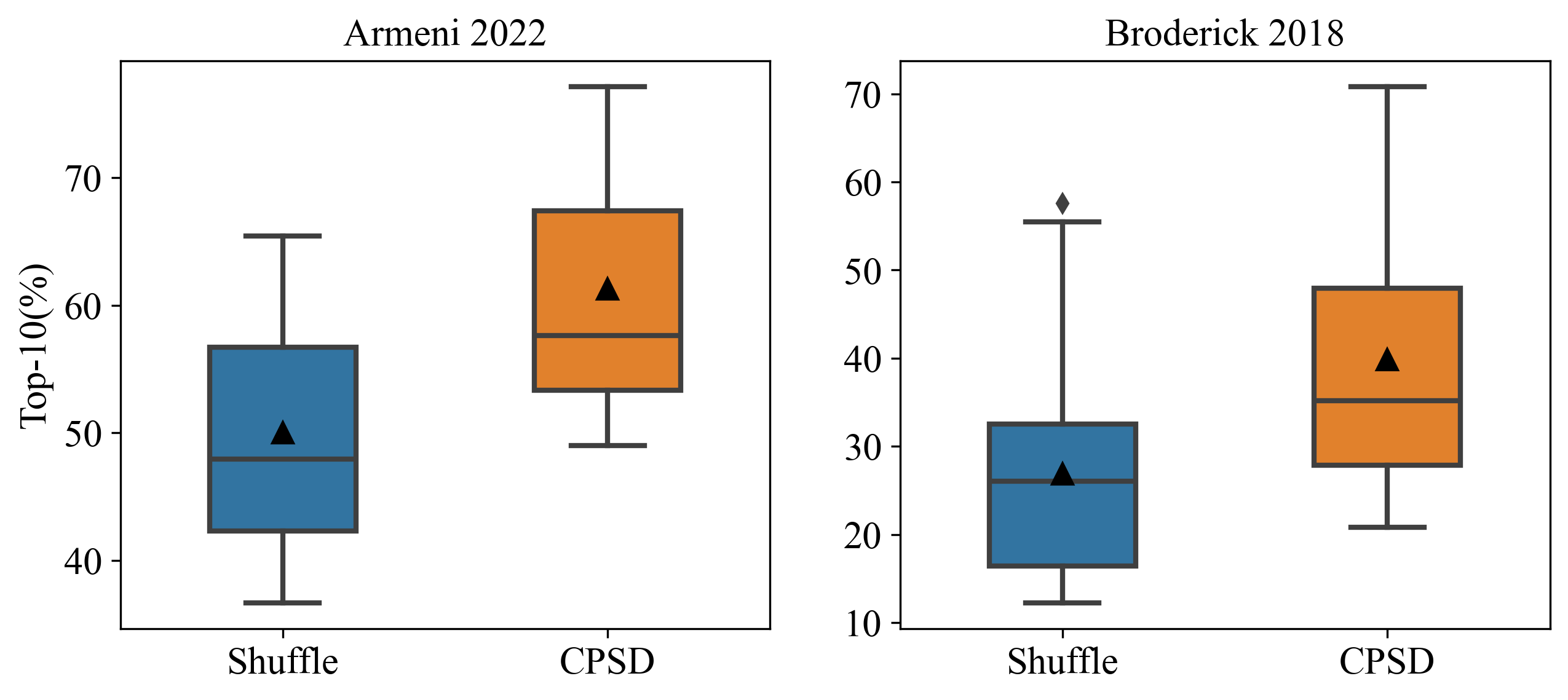} 
\caption{Verification of the validity of sensor locations as input. The performance improvement of CPSD compared to the Shuffle setting in both figures is statistically significant ($p<0.01$).}
\label{permute}
\end{figure}

To verify that the model effectively utilizes the spatial distribution information of the sensors, we randomly shuffled the sensor locations, then retrained and evaluated our framework using the same data. Due to computational constraints, we conducted experiments only on the Armeni 2022 and Broderick 2018 datasets. The results shown in Fig. \ref{permute} indicate that after shuffling, the model's Top-10 accuracy decreased by more than 11\% on both datasets. These results are statistically significant, demonstrating the model's effective use of sensor location information.

\begin{figure}[ht]
\centering
\includegraphics[width=8.8cm, height=3.1cm]{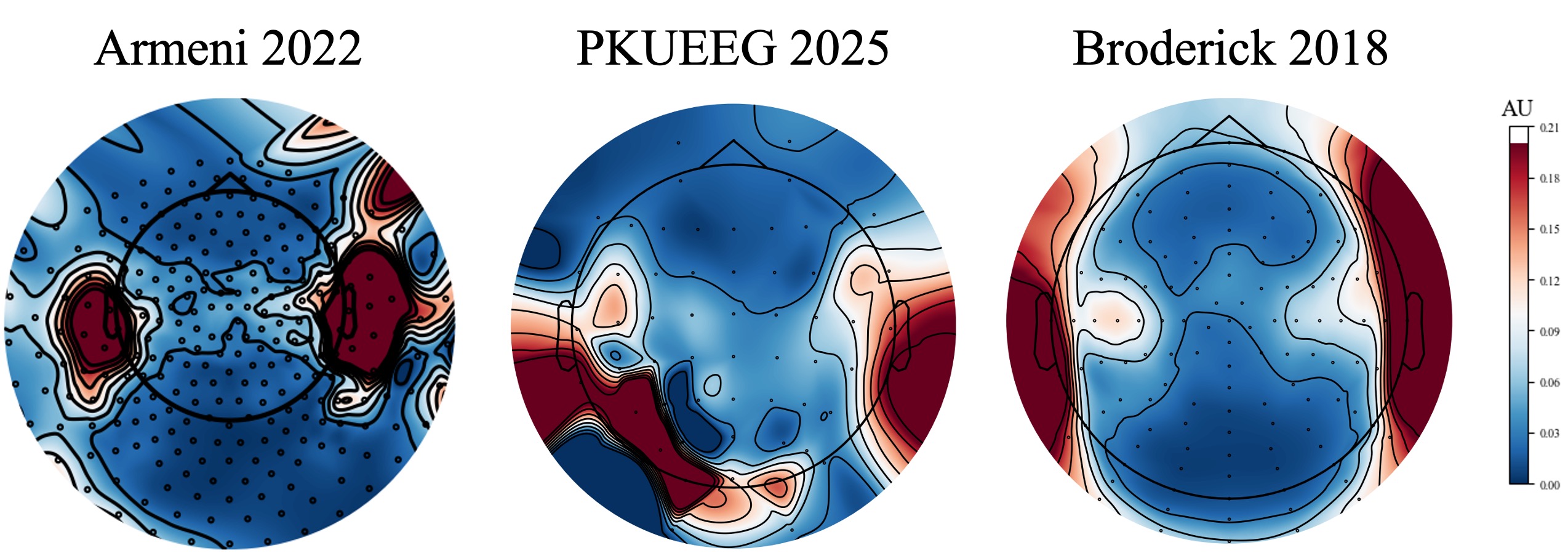} 
\caption{Visualization of channel weights from the PESA module across various datasets.}
\label{topo}
\end{figure}

We further interpret the output from the PESA module by visualizing the learned channel weights on MEG/EEG scalp topographies. Let ${\{W_i\}_{i\in S}}$ represent the PESA weights from models of different subjects. We first average these weights across subjects:

\begin{equation}
    W = \frac{1}{|S|}\sum_{i=1}^{|S|}W_{i}, W\in\mathbb{R}^{C\times D}
\end{equation} 

\noindent where $|S|$, $C$, and $D$ denote the number of subjects, input channels, and hidden dimensions, respectively. To highlight the contribution of different channels to the experimental results, we obtained the topography $T$ from Eq. \eqref{topograph}:

\begin{equation}
    T = \sqrt{W\odot WE}, T\in \mathbb{R}^{C} \label{topograph}
\end{equation}

\noindent where $\odot$ denotes the Hadamard product and $E=\frac{1}{D}(1,1,\ldots,1)^T \in \mathbb{R}^{D}$. The results for the three datasets are shown in Fig. \ref{topo}. For the perceived speech decoding task, the model assigns higher weights to sensors over the bilateral temporal regions, consistent with their role in auditory processing. These findings demonstrate that the PESA module effectively captures brain regions relevant to specific decoding tasks.

\section{Conclusion}
In this work, we propose a cross-subject perceived speech decoding method CPSD that employs a two-stage training approach: source model pre-training and personal specialization. In the source model pre-training stage, neural representations with high subject consistency are extracted through contrastive learning. During the personal specialization stage, consistent components are derived from the subject layer of the source model and adapted to individual differences using target subject data. To enhance the consistency of neural representations, this study introduces a positional encoding-based spatial attention module PESA. This module remaps MEG/EEG data to a standardized reference space to optimize the model’s generalizability and decoding performance. The performance of CPSD was evaluated on three perceived speech neural datasets encompassing different modalities and languages. Experimental results demonstrate that this method outperforms existing baseline models with statistically significant improvements. The effectiveness, robustness, and efficiency of the proposed method were validated through ablation study, hyperparameter search, comparisons with multi-subject setting, zero-shot decoding, neural representation consistency evaluation, sensor location validity analysis, and training time comparison. Finally, visualization of the PESA module’s channel weight matrix revealed higher weights near the bilateral temporal lobes, further confirming that the proposed method effectively focuses on critical brain regions associated with speech perception. 

\section{Future Work}
In future research, we will aim to improve the model's zero-shot decoding performance and explore additional applications. Potential solutions for extending the model to a broader range of applications include the following two points:

Studies have confirmed an overlap between the neural mechanisms underlying speech perception and speech production. For example, Hickok et al. demonstrated that Broca's area is involved both in motor planning for speech production and in recognizing and parsing speech signals during perception \cite{hickok2007cortical}. Additionally, there is a correlation between different speech production tasks: models trained on overt speech outperform those trained directly on imagined speech, indicating that the neural representations elicited by these two tasks overlap \cite{komeiji2024feasibility,proix2022imagined}. Therefore, developing a transfer learning framework that leverages different experimental settings may represent a promising technical approach for achieving non-invasive speech neuroprostheses.

Lastly, although brain foundation models have been extensively studied \cite{xiao2026brainomni,liu2026eeg,wang2025cbramod,jiang2024large}, their application to perceived speech decoding still faces significant limitations. These models rely heavily on tokenization techniques to achieve cross-device training \cite{yang2023biot}; however, the effectiveness of tokenizing perceived speech neural data remains insufficiently validated. Consequently, developing effective methods for cross-device decoding has become a critical challenge in implementing relevant foundation models. The PESA module proposed in this article decouples the input data dimensions from the model parameters, enabling control over the transformation of input data through channel position information. This approach allows a unified-dimensional neural representation to be fed into the encoder for further processing. Therefore, it is feasible to construct a brain foundation model based on the PESA module and evaluate the effectiveness of this method in speech decoding tasks.
 
\bibliographystyle{IEEEtran}
\bibliography{ref}

@String{Computer = "{IEEE} Computer" }

@article{wairagkar2025instantaneous,
  title={An instantaneous voice-synthesis neuroprosthesis},
  author={Wairagkar, Maitreyee and Card, Nicholas S and Singer-Clark, Tyler and Hou, Xianda and Iacobacci, Carrina and Miller, Lee M and Hochberg, Leigh R and Brandman, David M and Stavisky, Sergey D},
  journal={Nature},
  pages={1--8},
  year={2025},
  publisher={Nature Publishing Group UK London}
}

@article{card2024accurate,
  title={An accurate and rapidly calibrating speech neuroprosthesis},
  author={Card, Nicholas S and Wairagkar, Maitreyee and Iacobacci, Carrina and Hou, Xianda and Singer-Clark, Tyler and Willett, Francis R and Kunz, Erin M and Fan, Chaofei and Vahdati Nia, Maryam and Deo, Darrel R and others},
  journal={New England Journal of Medicine},
  volume={391},
  number={7},
  pages={609--618},
  year={2024},
  publisher={Mass Medical Soc}
}

@article{tang2023semantic,
  title={Semantic reconstruction of continuous language from non-invasive brain recordings},
  author={Tang, Jerry and LeBel, Amanda and Jain, Shailee and Huth, Alexander G},
  journal={Nature Neuroscience},
  volume={26},
  number={5},
  pages={858--866},
  year={2023},
  publisher={Nature Publishing Group US New York}
}

@article{zhang2025novel,
  title={A Novel Multimodal Method for Decoding Speech Perception from Brain Activities},
  author={Zhang, Aoke and Wang, Bo and Wu, Xihong and Chen, Jing},
  journal={ICASSP 2025-2025 IEEE International Conference on Acoustics, Speech and Signal Processing (ICASSP)},
  pages={1--5},
  year={2025},
}

@article{wang2024semantic,
  title={Semantic reconstruction of continuous language from meg signals},
  author={Wang, Bo and Xu, Xiran and Zhang, Longxiang and Xiao, Boda and Wu, Xihong and Chen, Jing},
  journal={ICASSP 2024-2024 IEEE International Conference on Acoustics, Speech and Signal Processing (ICASSP)},
  pages={2190--2194},
  year={2024},
}

@article{defossez2023decoding,
  title={Decoding speech perception from non-invasive brain recordings},
  author={D{\'e}fossez, Alexandre and Caucheteux, Charlotte and Rapin, J{\'e}r{\'e}my and Kabeli, Ori and King, Jean-R{\'e}mi},
  journal={Nature Machine Intelligence},
  volume={5},
  number={10},
  pages={1097--1107},
  year={2023},
  publisher={Nature Publishing Group UK London}
}

@article{xu2024convconcatnet,
  title={Convconcatnet: a deep convolutional neural network to reconstruct mel spectrogram from the eeg},
  author={Xu, Xiran and Wang, Bo and Yan, Yujie and Zhu, Haolin and Zhang, Zechen and Wu, Xihong and Chen, Jing},
  journal={2024 IEEE International Conference on Acoustics, Speech, and Signal Processing Workshops (ICASSPW)},
  pages={113--114},
  year={2024},
}

@article{accou2023decoding,
  title={Decoding of the speech envelope from EEG using the VLAAI deep neural network},
  author={Accou, Bernd and Vanthornhout, Jonas and hamme, Hugo Van and Francart, Tom},
  journal={Scientific Reports},
  volume={13},
  number={1},
  pages={812},
  year={2023},
  publisher={Nature Publishing Group UK London}
}

@article{myers2024individual,
  title={Individual differences in the perception of phonetic category structure predict speech-in-noise performance},
  author={Myers, Emily and Phillips, Matthew and Skoe, Erika},
  journal={The Journal of the Acoustical Society of America},
  volume={156},
  number={3},
  pages={1707--1719},
  year={2024},
  publisher={Acoustical Society of America}
}

@article{giovannone2021individual,
  title={Individual differences in lexical contributions to speech perception},
  author={Giovannone, Nikole and Theodore, Rachel M},
  journal={Journal of Speech, Language, and Hearing Research},
  volume={64},
  number={3},
  pages={707--724},
  year={2021},
  publisher={American Speech-Language-Hearing Association}
}

@article{dmochowski2012correlated,
  title={Correlated components of ongoing EEG point to emotionally laden attention--a possible marker of engagement?},
  author={Dmochowski, Jacek P and Sajda, Paul and Dias, Joao and Parra, Lucas C},
  journal={Frontiers in human neuroscience},
  volume={6},
  pages={112},
  year={2012},
  publisher={Frontiers Research Foundation}
}

@article{parra2018correlated,
  title={Correlated components analysis-extracting reliable dimensions in multivariate data},
  author={Parra, Lucas C and Haufe, Stefan and Dmochowski, Jacek P},
  journal={arXiv preprint arXiv:1801.08881},
  year={2018}
}

@article{radford2021learning,
  title={Learning transferable visual models from natural language supervision},
  author={Radford, Alec and Kim, Jong Wook and Hallacy, Chris and Ramesh, Aditya and Goh, Gabriel and Agarwal, Sandhini and Sastry, Girish and Askell, Amanda and Mishkin, Pamela and Clark, Jack and others},
  journal={International conference on machine learning},
  pages={8748--8763},
  year={2021},
}

@article{hendrycks2016gaussian,
  title={Gaussian error linear units (gelus)},
  author={Hendrycks, Dan and Gimpel, Kevin},
  journal={arXiv preprint arXiv:1606.08415},
  year={2016}
}

@article{ioffe2015batch,
  title={Batch normalization: Accelerating deep network training by reducing internal covariate shift},
  author={Ioffe, Sergey and Szegedy, Christian},
  journal={International conference on machine learning},
  pages={448--456},
  year={2015},
}

@article{dauphin2017language,
  title={Language modeling with gated convolutional networks},
  author={Dauphin, Yann N and Fan, Angela and Auli, Michael and Grangier, David},
  journal={International conference on machine learning},
  pages={933--941},
  year={2017},
}

@article{c:22,
      title={Attention Is All You Need}, 
      author={Ashish Vaswani and Noam Shazeer and Niki Parmar and Jakob Uszkoreit and Llion Jones and Aidan N. Gomez and Lukasz Kaiser and Illia Polosukhin},
      year={2017},
      journal={arXiv preprint arXiv:1706.03762},
}

@article{ba2016layer,
  title={Layer normalization},
  author={Ba, Jimmy Lei and Kiros, Jamie Ryan and Hinton, Geoffrey E},
  journal={arXiv preprint arXiv:1607.06450},
  year={2016}
}

@article{jiang2024large,
  title={Large brain model for learning generic representations with tremendous EEG data in BCI},
  author={Jiang, Wei-Bang and Zhao, Liming and Lu, Bao-Liang},
  journal={International Conference on Learning Representations},
  volume={2024},
  pages={16405--16426},
  year={2024}
}

@article{akbari2019towards,
  title={Towards reconstructing intelligible speech from the human auditory cortex},
  author={Akbari, Hassan and Khalighinejad, Bahar and Herrero, Jose L and Mehta, Ashesh D and Mesgarani, Nima},
  journal={Scientific reports},
  volume={9},
  number={1},
  pages={874},
  year={2019},
  publisher={Nature Publishing Group UK London}
}

@article{kraemer2005sound,
  title={Sound of silence activates auditory cortex},
  author={Kraemer, David JM and Macrae, C Neil and Green, Adam E and Kelley, William M},
  journal={Nature},
  volume={434},
  number={7030},
  pages={158--158},
  year={2005},
  publisher={Nature Publishing Group UK London}
}

@article{wheeler2000memory,
  title={Memory's echo: vivid remembering reactivates sensory-specific cortex},
  author={Wheeler, Mark E and Petersen, Steven E and Buckner, Randy L},
  journal={Proceedings of the National Academy of Sciences},
  volume={97},
  number={20},
  pages={11125--11129},
  year={2000},
  publisher={The National Academy of Sciences}
}

@article{li2024visual,
  title={Visual decoding and reconstruction via EEG embeddings with guided diffusion},
  author={Li, Dongyang and Wei, Chen and Li, Shiying and Zou, Jiachen and Liu, Quanying},
  journal={Proceedings of the 38th International Conference on Neural Information Processing Systems},
  pages={102822--102864},
  year={2024}
}

@misc{nastase2019measuring,
  title={Measuring shared responses across subjects using intersubject correlation},
  author={Nastase, Samuel A and Gazzola, Valeria and Hasson, Uri and Keysers, Christian},
  journal={Social cognitive and affective neuroscience},
  volume={14},
  number={6},
  pages={667--685},
  year={2019},
  publisher={Oxford University Press}
}

@article{simanova2010identifying,
  title={Identifying object categories from event-related EEG: toward decoding of conceptual representations},
  author={Simanova, Irina and Van Gerven, Marcel and Oostenveld, Robert and Hagoort, Peter},
  journal={PloS one},
  volume={5},
  number={12},
  pages={e14465},
  year={2010},
  publisher={Public Library of Science San Francisco, USA}
}

@article{zhang2022self,
  title={Self-supervised contrastive pre-training for time series via time-frequency consistency},
  author={Zhang, Xiang and Zhao, Ziyuan and Tsiligkaridis, Theodoros and Zitnik, Marinka},
  journal={Advances in neural information processing systems},
  volume={35},
  pages={3988--4003},
  year={2022}
}

@article{dong2023simmtm,
  title={Simmtm: A simple pre-training framework for masked time-series modeling},
  author={Dong, Jiaxiang and Wu, Haixu and Zhang, Haoran and Zhang, Li and Wang, Jianmin and Long, Mingsheng},
  journal={Advances in Neural Information Processing Systems},
  volume={36},
  pages={29996--30025},
  year={2023}
}

@article{yang2023biot,
  title={Biot: Biosignal transformer for cross-data learning in the wild},
  author={Yang, Chaoqi and Westover, M and Sun, Jimeng},
  journal={Advances in Neural Information Processing Systems},
  volume={36},
  pages={78240--78260},
  year={2023}
}

@article{bethge2022domain,
  title={Domain-invariant representation learning from EEG with private encoders},
  author={Bethge, David and Hallgarten, Philipp and Grosse-Puppendahl, Tobias and Kari, Mohamed and Mikut, Ralf and Schmidt, Albrecht and {\"O}zdenizci, Ozan},
  journal={ICASSP 2022-2022 IEEE International Conference on Acoustics, Speech and Signal Processing (ICASSP)},
  pages={1236--1240},
  year={2022},
}

@article{wang2024dmmr,
  title={DMMR: Cross-subject domain generalization for EEG-based emotion recognition via denoising mixed mutual reconstruction},
  author={Wang, Yiming and Zhang, Bin and Tang, Yujiao},
  journal={Proceedings of the AAAI conference on artificial intelligence},
  volume={38},
  number={1},
  pages={628--636},
  year={2024}
}

@article{ma2019reducing,
  title={Reducing the subject variability of EEG signals with adversarial domain generalization},
  author={Ma, Bo-Qun and Li, He and Zheng, Wei-Long and Lu, Bao-Liang},
  journal={International Conference on Neural Information Processing},
  pages={30--42},
  year={2019},
}

@article{shen2024contrastive,
  title={Contrastive learning of shared spatiotemporal EEG representations across individuals for naturalistic neuroscience},
  author={Shen, Xinke and Tao, Lingyi and Chen, Xuyang and Song, Sen and Liu, Quanying and Zhang, Dan},
  journal={NeuroImage},
  volume={301},
  pages={120890},
  year={2024},
  publisher={Elsevier}
}

@article{broderick2018electrophysiological,
  title={Electrophysiological correlates of semantic dissimilarity reflect the comprehension of natural, narrative speech},
  author={Broderick, Michael P and Anderson, Andrew J and Di Liberto, Giovanni M and Crosse, Michael J and Lalor, Edmund C},
  journal={Current Biology},
  volume={28},
  number={5},
  pages={803--809},
  year={2018},
  publisher={Elsevier}
}

@article{armeni202210,
  title={A 10-hour within-participant magnetoencephalography narrative dataset to test models of language comprehension},
  author={Armeni, Kristijan and G{\"u}{\c{c}}l{\"u}, Umut and van Gerven, Marcel and Schoffelen, Jan-Mathijs},
  journal={Scientific Data},
  volume={9},
  number={1},
  pages={278},
  year={2022},
  publisher={Nature Publishing Group UK London}
}

@article{hamalainen1993magnetoencephalography,
  title={Magnetoencephalography—theory, instrumentation, and applications to noninvasive studies of the working human brain},
  author={H{\"a}m{\"a}l{\"a}inen, Matti and Hari, Riitta and Ilmoniemi, Risto J and Knuutila, Jukka and Lounasmaa, Olli V},
  journal={Reviews of modern Physics},
  volume={65},
  number={2},
  pages={413},
  year={1993},
  publisher={APS}
}

@article{baevski2020wav2vec,
  title={wav2vec 2.0: A framework for self-supervised learning of speech representations},
  author={Baevski, Alexei and Zhou, Yuhao and Mohamed, Abdelrahman and Auli, Michael},
  journal={Advances in neural information processing systems},
  volume={33},
  pages={12449--12460},
  year={2020}
}

@article{liuitransformer,
  title={iTransformer: Inverted Transformers Are Effective for Time Series Forecasting},
  author={Liu, Yong and Hu, Tengge and Zhang, Haoran and Wu, Haixu and Wang, Shiyu and Ma, Lintao and Long, Mingsheng},
  journal={The Twelfth International Conference on Learning Representations},
  year={2024},
}

@article{pereira2018toward,
  title={Toward a universal decoder of linguistic meaning from brain activation},
  author={Pereira, Francisco and Lou, Bin and Pritchett, Brianna and Ritter, Samuel and Gershman, Samuel J and Kanwisher, Nancy and Botvinick, Matthew and Fedorenko, Evelina},
  journal={Nature communications},
  volume={9},
  number={1},
  pages={963},
  year={2018},
  publisher={Nature Publishing Group UK London}
}

@article{hu2025cross,
  title={Cross-subject emotion recognition with contrastive learning based on EEG signal correlations},
  author={Hu, Mengting and Xu, Dan and He, Kangjian and Zhao, Kunyuan and Zhang, Hao},
  journal={Biomedical Signal Processing and Control},
  volume={104},
  pages={107511},
  year={2025},
  publisher={Elsevier}
}

@article{hasson2004intersubject,
  title={Intersubject synchronization of cortical activity during natural vision},
  author={Hasson, Uri and Nir, Yuval and Levy, Ifat and Fuhrmann, Galit and Malach, Rafael},
  journal={science},
  volume={303},
  number={5664},
  pages={1634--1640},
  year={2004},
  publisher={American Association for the Advancement of Science}
}

@article{gramfort2013meg,
  title={MEG and EEG data analysis with MNE-Python},
  author={Gramfort, Alexandre and Luessi, Martin and Larson, Eric and Engemann, Denis A and Strohmeier, Daniel and Brodbeck, Christian and Goj, Roman and Jas, Mainak and Brooks, Teon and Parkkonen, Lauri and others},
  journal={Frontiers in Neuroinformatics},
  volume={7},
  pages={267},
  year={2013},
  publisher={Frontiers Media SA}
}

@article{levy2026noninvasive,
  title={Noninvasive decoding of typed sentences from human brain activity},
  author={L{\'e}vy, Jarod and Zhang, Mingfang and Pinet, Svetlana and Rapin, J{\'e}r{\'e}my and Banville, Hubert and d’Ascoli, St{\'e}phane and King, Jean-R{\'e}mi},
  journal={Nature Neuroscience},
  pages={1--7},
  year={2026},
  publisher={Nature Publishing Group US New York}
}

@article{bhattacharjee2026aligning,
  title={Aligning brains into a shared space improves their alignment with large language models},
  author={Bhattacharjee, Arnab and Zada, Zaid and Wang, Haocheng and Aubrey, Bobbi and Doyle, Werner and Dugan, Patricia and Friedman, Daniel and Devinsky, Orrin and Flinker, Adeen and Ramadge, Peter J and others},
  journal={Nature Computational Science},
  volume={6},
  number={2},
  pages={169--178},
  year={2026},
  publisher={Nature Publishing Group US New York}
}

@article{zou2026constituent,
  title={Constituent-constrained word prediction during language comprehension},
  author={Zou, Jiajie and Poeppel, David and Ding, Nai},
  journal={Nature Neuroscience},
  volume={29},
  pages={1498--1509},
  year={2026},
  publisher={Nature Publishing Group US New York}
}

@article{silva2024speech,
  title={The speech neuroprosthesis},
  author={Silva, Alexander B and Littlejohn, Kaylo T and Liu, Jessie R and Moses, David A and Chang, Edward F},
  journal={Nature Reviews Neuroscience},
  volume={25},
  number={7},
  pages={473--492},
  year={2024},
  publisher={Nature Publishing Group UK London}
}

@article{hickok2007cortical,
  title={The cortical organization of speech processing},
  author={Hickok, Gregory and Poeppel, David},
  journal={Nature reviews neuroscience},
  volume={8},
  number={5},
  pages={393--402},
  year={2007},
  publisher={Nature Publishing Group UK London}
}

@article{skipper2005listening,
  title={Listening to talking faces: motor cortical activation during speech perception},
  author={Skipper, Jeremy I and Nusbaum, Howard C and Small, Steven L},
  journal={Neuroimage},
  volume={25},
  number={1},
  pages={76--89},
  year={2005},
  publisher={Elsevier}
}

@article{benchetrit2024brain,
  title={Brain decoding: toward real-time reconstruction of visual perception},
  author={Benchetrit, Yohann and Banville, Hubert and King, Jean-R{\'e}mi},
  journal={International Conference on Learning Representations},
  volume={2024},
  pages={7846--7858},
  year={2024}
}

@article{li2023blip,
  title={Blip-2: Bootstrapping language-image pre-training with frozen image encoders and large language models},
  author={Li, Junnan and Li, Dongxu and Savarese, Silvio and Hoi, Steven},
  journal={International conference on machine learning},
  pages={19730--19742},
  year={2023},
}

@article{mu2022slip,
  title={Slip: Self-supervision meets language-image pre-training},
  author={Mu, Norman and Kirillov, Alexander and Wagner, David and Xie, Saining},
  journal={European conference on computer vision},
  pages={529--544},
  year={2022},
}

@article{polikov2005response,
  title={Response of brain tissue to chronically implanted neural electrodes},
  author={Polikov, Vadim S. and Tresco, Patrick A. and Reichert, William M.},
  journal={Journal of Neuroscience Methods},
  volume={148},
  number={1},
  pages={1--18},
  year={2005},
  doi={10.1016/j.jneumeth.2005.08.015},
  pmid={16198003}
}

@article{yang2025neural,
  title={Neural electrodes for brain-computer interface system: From rigid to soft},
  author={Yang, Dan and Tian, Gongwei and Chen, Jianhui and Liu, Yan and Fatima, Esha and Qiu, Jichuan and Malek, Nik Ahmad Nizam Nik and Qi, Dianpeng},
  journal={BMEMat},
  volume={3},
  number={3},
  pages={e12130},
  year={2025},
  publisher={Wiley Online Library}
}

@article{chen2025long,
  title={Long-Term Stable Subdural Recordings Enabled by Fibrosis-Resistant Hydrogel-Integrated $\mu$ECoG Arrays},
  author={Chen, Lin and Zhong, Hao and Wang, Linghao and Xu, Liju and Fan, Wanying and Zhao, Yongpeng and Zhang, Huiling and Shen, Yang and Wu, Kai and Fu, Xin and others},
  journal={Advanced Science},
  volume={12},
  number={47},
  pages={e15453},
  year={2025},
  publisher={Wiley Online Library}
}

@article{Sarvas1987,
  title = {Basic mathematical and electromagnetic concepts of the biomagnetic inverse problem},
  author = {Sarvas, Jukka},
  journal = {Physics in Medicine \& Biology},
  year = {1987},
  number = {1},
  pages = {11--22},
  volume = {32},
}

@article{sanei2013eeg,
  title={EEG signal processing},
  author={Sanei, Saeid and Chambers, Jonathon A},
  year={2013},
  journal={John Wiley \& Sons}
}

@article{lin2006distributed,
  title={Distributed current estimates using cortical orientation constraints},
  author={Lin, Fa-Hsuan and Belliveau, John W and Dale, Anders M and H{\"a}m{\"a}l{\"a}inen, Matti S},
  journal={Human brain mapping},
  volume={27},
  number={1},
  pages={1--13},
  year={2006},
  publisher={Wiley Online Library}
}

@article{lu2026brain,
  title={Brain-Inspired fMRI-to-Text Decoding via Incremental and Wrap-Up Language Modeling},
  author={Lu, Wentao and Nie, Dong and Xue, Pengcheng and Cui, Zheng and Li, Piji and Zhang, Daoqiang and Wen, Xuyun},
  journal={Advances in Neural Information Processing Systems},
  volume={38},
  pages={149986--150009},
  year={2026}
}

@article{chen2024open,
  title={Open-vocabulary auditory neural decoding using fMRI-prompted llm},
  author={Chen, Xiaoyu and Du, Changde and Liu, Che and Wang, Yizhe and He, Huiguang},
  journal={arXiv preprint arXiv:2405.07840},
  year={2024}
}

@article{wang2026hierarchical,
  title={Hierarchical Decoding of Perceived Speech From Non-Invasive Brain Recordings},
  author={Wang, Bo and Xu, Xiran and Xiao, Boda and Zheng, Linze and Wu, Xihong and Cheng, Heping and Chen, Jing},
  journal={IEEE Transactions on Neural Systems and Rehabilitation Engineering},
  volume={34},
  pages={2349--2360},
  year={2026},
  publisher={IEEE}
}

@article{huth2016natural,
  title={Natural speech reveals the semantic maps that tile human cerebral cortex},
  author={Huth, Alexander G and De Heer, Wendy A and Griffiths, Thomas L and Theunissen, Fr{\'e}d{\'e}ric E and Gallant, Jack L},
  journal={Nature},
  volume={532},
  number={7600},
  pages={453--458},
  year={2016},
  publisher={Nature Publishing Group UK London}
}

@article{li2020perils,
  title={The perils and pitfalls of block design for EEG classification experiments},
  author={Li, Ren and Johansen, Jared S and Ahmed, Hamad and Ilyevsky, Thomas V and Wilbur, Ronnie B and Bharadwaj, Hari M and Siskind, Jeffrey Mark},
  journal={IEEE Transactions on Pattern Analysis and Machine Intelligence},
  volume={43},
  number={1},
  pages={316--333},
  year={2020},
  publisher={IEEE}
}

@article{xu2026impacts,
  title={The impacts of temporal autocorrelations on EEG decoding},
  author={Xu, Xiran and Wang, Bo and Xiao, Boda and Niu, Yadong and Wang, Yiwen and Wu, Xihong and Cheng, Heping and Chen, Jing},
  journal={Biomedical Signal Processing and Control},
  volume={113},
  pages={108783},
  year={2026},
  publisher={Elsevier}
}

@article{xiao2026brainomni,
  title={Brainomni: A brain foundation model for unified eeg and meg signals},
  author={Xiao, Qinfan and Cui, Ziyun and Zhang, Chi and Chen, Siqi and Wu, Wen and Thwaites, Andrew and Woolgar, Alexandra and Zhou, Bowen and Zhang, Chao},
  journal={Advances in Neural Information Processing Systems},
  volume={38},
  pages={41179--41212},
  year={2026}
}

@article{wang2025cbramod,
  title={Cbramod: A criss-cross brain foundation model for eeg decoding},
  author={Wang, Jiquan and Zhao, Sha and Luo, Zhiling and Zhou, Yangxuan and Jiang, Haiteng and Li, Shijian and Li, Tao and Pan, Gang},
  journal={International conference on learning representations},
  volume={2025},
  pages={75310--75346},
  year={2025}
}

@article{komeiji2024feasibility,
  title={Feasibility of decoding covert speech in ECoG with a Transformer trained on overt speech},
  author={Komeiji, Shuji and Mitsuhashi, Takumi and Iimura, Yasushi and Suzuki, Hiroharu and Sugano, Hidenori and Shinoda, Koichi and Tanaka, Toshihisa},
  journal={Scientific Reports},
  volume={14},
  number={1},
  pages={11491},
  year={2024},
  publisher={Nature Publishing Group UK London}
}

@article{proix2022imagined,
  title={Imagined speech can be decoded from low-and cross-frequency intracranial EEG features},
  author={Proix, Timoth{\'e}e and Delgado Saa, Jaime and Christen, Andy and Martin, Stephanie and Pasley, Brian N and Knight, Robert T and Tian, Xing and Poeppel, David and Doyle, Werner K and Devinsky, Orrin and others},
  journal={Nature communications},
  volume={13},
  number={1},
  pages={48},
  year={2022},
  publisher={Nature Publishing Group UK London}
}

@article{leuthardt2021defining,
  title={Defining surgical terminology and risk for brain computer interface technologies},
  author={Leuthardt, Eric C and Moran, Daniel W and Mullen, Tim R},
  journal={Frontiers in Neuroscience},
  volume={15},
  pages={599549},
  year={2021},
  publisher={Frontiers}
}

@article{peksa2023state,
  title={State-of-the-art on brain-computer interface technology},
  author={Peksa, Janis and Mamchur, Dmytro},
  journal={Sensors},
  volume={23},
  number={13},
  pages={6001},
  year={2023},
  publisher={MDPI}
}

@article{liu2026eeg,
  title={EEG Foundation Models: Progresses, Benchmarking, and Open Problems},
  author={Liu, Dingkun and Chen, Yuheng and Chen, Zhu and Cui, Zhenyao and Wen, Yaozhi and An, Jiayu and Luo, Jingwei and Wu, Dongrui},
  journal={arXiv preprint arXiv:2601.17883},
  year={2026}
}

@article{kingma2014adam,
  title={Adam: A method for stochastic optimization},
  author={Kingma, Diederik P and Ba, Jimmy},
  journal={arXiv preprint arXiv:1412.6980},
  year={2014}
}

@article{jung1998independent,
  title={Independent component analysis of electroencephalographic and event-related potential data},
  author={Jung, Tzyy-Ping and Makeig, Scott and Bell, Anthony J and Sejnowski, Terrence J},
  journal={Central auditory processing and neural modeling},
  pages={189--197},
  year={1998},
}


\vfill

\end{document}